\documentclass[useAMS,usenatbib,usegraphics]{mn2e}
\usepackage{ifpdf} 
\pdfoutput=1
\usepackage{graphicx,tablefootnote}
\usepackage{latexsym,amssymb}
\usepackage{amsmath}
\usepackage{gensymb}
\usepackage{capt-of}
\usepackage{textpos}
\usepackage{url}
\usepackage{multirow}
\usepackage{color}
\usepackage[T1]{fontenc}
\usepackage{longtable}
\usepackage{hyperref}
\usepackage{float}
\usepackage{siunitx}
\usepackage{textcomp}

\hypersetup{
    bookmarks=true,         % show bookmarks bar?
    unicode=false,          % non-Latin characters in Acrobat�s bookmarks
    pdftoolbar=true,        % show Acrobat�s toolbar?
    pdfmenubar=true,        % show Acrobat�s menu?
    pdffitwindow=false,     % window fit to page when opened
    pdfstartview={FitH},    % fits the width of the page to the window
    pdftitle={My title},    % title
    pdfauthor={Author},     % author
    pdfsubject={Subject},   % subject of the document
    pdfcreator={Creator},   % creator of the document
    pdfproducer={Producer}, % producer of the document
    pdfkeywords={keyword1} {key2} {key3}, % list of keywords
    pdfnewwindow=true,      % links in new window
    colorlinks=true,       % false: boxed links; true: colored links
    linkcolor=black,          % color of internal links (change box color with linkbordercolor)
    citecolor=black,        % color of links to bibliography
    filecolor=magenta,      % color of file links
    urlcolor=black,           % color of external links
    linkbordercolor={1 0 0},
    citebordercolor={0 1 0}
}

\makeatletter
\Hy@AtBeginDocument{%
  \def\@pdfborder{0 0 1}% Overrides border definition set with colorlinks=true
  \def\@pdfborderstyle{/S/U/W 0}% Overrides border style set with colorlinks=true
}
\makeatother

\newcommand{\kms}{km\,s$^{-1}$}

\newcommand{\Msun}{M$_\odot$}
\newcommand{\Lsun}{L$_\odot$}

\newcommand{\kmsM}{km\,s$^{-1}$\,Mpc$^{-1}$}

\newcommand{\ML}{\ensuremath{\Upsilon}}

\newcommand{\MLstar}{\ensuremath{\Upsilon_\star}}

\newcommand{\multi}{\textsc{Multidrizzle}}  %Project Multidrizzle
\newcommand{\holi}{H0LiCOW}			     	%Project H0LiCOW
\newcommand{\hst}{\textit{HST}}             %HST
\newcommand{\jwst}{\textit{JWST}}           %JWST
\newcommand{\tmt}{\textit{TMT}}             %TMT
\newcommand{\elt}{\textit{E-ELT}}           %E-ELT
\newcommand{\sersic}{S\'{e}rsic}		    %S�rsic
\newcommand{\lcdm}{$\Lambda$CDM}			%LCDM
\newcommand{\rxj}{RXJ1131$-$1231}             %WMAP
\newcommand{\h}{$H_\mathrm{0}$}				%H0
\newcommand{\kext}{\ensuremath{\kappa_\mathrm{ext}}}			%kext
\newcommand{\dt}{\ensuremath{D_{\Delta t}}}						%Ddt
\newcommand{\dtmod}{\ensuremath{D_{\Delta t}^{\rm model}}} 		%Ddt^mod
\newcommand{\dd}{\ensuremath{D_\mathrm{d}}}    					%Dd
\newcommand{\ddmod}{\ensuremath{D_\mathrm{d}^{\rm model}}}		%Dd^mod
\newcommand{\ds}{\ensuremath{D_\mathrm{s}}}    					%Ds
\newcommand{\dds}{\ensuremath{D_\mathrm{ds}}}		    		%Dds
\newcommand{\zd}{\ensuremath{z_\mathrm{d}}}    					%Dds
\newcommand{\vrms}{\ensuremath{V_\mathrm{rms}}}         		%Vrms
\newcommand{\vlos}{\ensuremath{\overline{v_{\mathrm{LOS}}^{2}}}}%Vlos

\title[Cosmographic forecasts with strong lensing and JWST stellar kinematics]{Time-delay cosmographic forecasts with strong lensing and JWST stellar kinematics}
\author[A. Y{\i}ld{\i}r{\i}m, S.~H.~Suyu and A.~Halkola]
{\parbox{\textwidth}{Ak{\i}n Y{\i}ld{\i}r{\i}m$^{1}$\thanks{E-mail: yildirim@mpa-garching.mpg.de},
Sherry H.~Suyu$^{1, 2, 3}$, Aleksi Halkola$^{4}$}\vspace{0.4cm}\\
\parbox{\textwidth}{
$^{1}$Max Planck Institute for Astrophysics, Karl-Schwarzschild-Str. 1, 85748 Garching, Germany\\
$^{2}$Physik-Department, Technische Universit\"at M\"unchen, James-Franck-Str. 1, 85748 Garching, Germany\\
$^{3}$Institute of Astronomy and Astrophysics, Academia Sinica, 11F of ASMAB, No.1, Section 4, Roosevelt Road, Taipei 10617, Taiwan\\
$^{4}$Py\"orrekuja 5 A, 04300 Tuusula, Finland }}
\begin{document}

\date{Accepted for publication in MNRAS}
%\date{Accepted 1988 December 15. Received 1988 December 14; in original form 1988 October 11}

%\pagerange{\pageref{firstpage}--\pageref{lastpage}} \pubyear{2002}

\def\LaTeX{L\kern-.36em\raise.3ex\hbox{a}\kern-.15em
    T\kern-.1667em\lower.7ex\hbox{E}\kern-.125emX}

\maketitle

\label{firstpage}

\begin{abstract}

We present a joint strong lensing and stellar dynamical framework for future time-delay cosmography purposes. Based on a pixelated source reconstruction and the axisymmetric Jeans equations, we are capable of constraining cosmological distances and hence the current expansion rate of the Universe (\h) to the few percent level per lens, when high signal-to-noise integral field unit (IFU) observations from the next generation of telescopes become available. For illustrating the power of this method, we mock up IFU stellar kinematic data of the prominent lens system \rxj, given the specifications of the James Webb Space Telescope. Our analysis shows that the time-delay distance (\dt) can be constrained with 3.1\% uncertainty at best, if future IFU stellar kinematics are included in the fit and if the set of candidate model parameterisations contains the true lens potential. These constraints would translate to a 3.2\% precision measurement on \h\ in flat \lcdm\ cosmology from the single lens \rxj, and can be expected to yield an \h\ measure with $\le$ 2.0\% uncertainty, if similar gains in precision can be reached for two additional lens systems. Moreover, the angular diameter distance (\dd) to \rxj\ can be constrained with 2.4\% precision, providing two distance measurements from a single lens system, which is extremely powerful to further constrain the matter density ($\Omega_{\mathrm{m}}$). The measurement accuracy of \dd, however, is highly sensitive to any systematics in the measurement of the stellar kinematics. For both distance measurements, we strongly advise to probe a large set of physically motivated lens potentials in the future, to minimise the systematic errors associated with the lens mass parameterisation.

\end{abstract}


%============================= Section 1 =============================
\section{Introduction}
\label{sec:introduction}
%=====================================================================

According to our standard cosmological model, we live in a flat, cold, dark matter and dark energy dominated Universe ($\Lambda$CDM). While little is yet known about the nature of dark matter and dark energy, our standard cosmological model - anchored mainly through measurements of anisotropies in the Cosmic Microwave Background \citep[CMB;][]{2000Natur.404..955D,2000ApJ...545L...5H,2003ApJS..148....1B} - has been well established and provides an accurate description of e.g. the
large scale structure formation and distribution \citep[see;][for a review]{2006Natur.440.1137S} and the abundance of various elements in the Universe. As powerful as this model is, however, it has been constantly facing challenges. On small scales, the well-known "core-cusp" issue \citep{1994Natur.370..629M,1998ApJ...499...41M} as
well as the "Missing Satellites Problem" \citep{1993MNRAS.264..201K,1999ApJ...522...82K,1999ApJ...524L..19M,2011MNRAS.415L..40B} are stubbornly defying predictions from cosmological \textit{N}-body
simulations within the \lcdm\ framework. Certainly, some of the disagreements can be attributed to observational effects \citep{2000AJ....119.1579V}, underlying modelling assumptions \citep{2009MNRAS.393L..50E} and our ignorance of the small-scale physics and baryonic feedback processes and interactions \citep{2011AJ....142...24O}. But, while more recent studies claim to solve some of the small scale discrepancies, by thoroughly accounting for e.g. star formation, supernovae feedback \citep[e.g.][]{2012ApJ...761...71Z,2014ApJ...789L..17M,2016MNRAS.456.3542T} and environmental impact such as ram pressure and tidal stripping \citep[e.g.][]{2013ApJ...765...22B,2016MNRAS.457L..74D,2016MNRAS.457.1931S}, similar discrepancies between theory and observations also arise at large scales. After making assumptions for a handful of parameters, such as spatial flatness and a constant dark energy equation of state of $-1$ (corresponding to the cosmological constant $\Lambda$), the standard cosmological model provides stringent constraints for the expansion rate of the Universe \h\ through observations of the CMB \citep{2018arXiv180706209P}, which appears to be at odds with local measurements based on Cepheids and Type Ia Supernovae \citep{2018ApJ...861..126R,2018ApJ...855..136R,2019ApJ...876...85R}. Especially the latter has been of particular interest lately. Given the significant 4.4$\sigma$ (i.e. $\sim$9.8\%) discrepancy between the most recent measurements from both the Planck Collaboration and the Cepheid distance ladder, this result is either interpreted as corroborating evidence for a non-standard cosmological model \citep[e.g.][]{2019A&A...628L...4L} or claimed to be part of unknown systematic effects that are not properly accounted for \citep[e.g.][]{2019MNRAS.484L..64S}. Naturally, relaxing our assumption about spatial flatness, a constant dark energy equation of state (i.e., not fixed to $-1$ that corresponds to the cosmological constant $\Lambda$) or increasing the number of relativistic species would allow us to reconcile both measurements. But, before such drastic conclusions are drawn, independent measurements of \h\ should be carried out, if feasible, to test for possibly unknown systematics in any single method and to assess the need for physics beyond the standard model.

Time-delay cosmography \citep[TDC;][]{1964MNRAS.128..307R} provides a methodologically independent tool for measuring \h\ to the percent level \citep[see recent reviews by, e.g.,][]{2016A&ARv..24...11T, 2018SSRv..214...91S}. By means of a multiply imaged, time-variable background source and an accurate description of the foreground lens mass distribution, the time-delay distance (\dt) can be inferred, which is inversely proportional to the Hubble-Lema\^{i}tre constant \h. The technique has long been plagued by poor time-delay measurements, invalid assumptions about the lens mass profile and systematic errors. However, it has been demonstrated that exhaustive studies of lensed quasars with exquisite light curves allow the measurement of \h\ for a single system with an accuracy of $\sim$ 7\% \citep{2010ApJ...711..201S,2013ApJ...766...70S}. In addition, it was shown that TDC leads to tight constraints on other cosmological parameters, competing with those from contemporary Baryon Acoustic Peak studies, when each probe is combined with the CMB \citep{2013ApJ...766...70S}.

In light of the aforementioned discrepancy between the current best cosmological probes, TDC has gained momentum and the \holi\footnote{\h\ Lenses in COSMOGRAIL's Wellspring; \url{http://www.h0licow.org/}}\ (\h\ Lenses in COSMOGRAIL's Wellspring) program has been initiated, which aims at measuring \h\ with better than 3.5\% precision and accuracy \citep{2017MNRAS.468.2590S}. As part of these efforts, complementary data sets consisting of i) high-cadence and long-baseline monitoring of quasar light curves, mostly through COSMOGRAIL\footnote{COSmological MOnitoring of GRAvItational Lenses; \url{https://cosmograil.epfl.ch}}, ii) high-spatially resolved photometric observations of the foreground lenses, lensed background quasars and quasar hosts, iii) wide-field photometric and spectroscopic observations of the lens' environments and iv) stellar kinematic data of the foreground lenses have been obtained. Each of these ingredients are crucial to break the inherent modelling degeneracies, i.e. the mass-sheet degeneracy \citep[MSD;][]{1985ApJ...289L...1F}, in strong lensing studies and to reliably pin down the time-delay distance and hence the Hubble-Lema\^itre constant in a single system. As of now, H0LiCOW reports a 3.0\% measurement of \h\ in flat LCDM, based on a joint analysis of 6 strong lensing systems \citep{2019arXiv190704869W,2019arXiv190702533C,2019arXiv190509338R,2019MNRAS.484.4726B,2017MNRAS.465.4914B,2017MNRAS.465.4895W,2014ApJ...788L..35S,2010ApJ...711..201S}. Yet, it is noteworthy that an \h\ measurement which is comparable in precision with the best available probes (i.e. $\sim$ 2\%) would still require the combination of almost a dozen lenses \citep{2018MNRAS.473..210S}. Accordingly, a measurement with 1\% precision - a value that is considered as being highly beneficial for any Stage III and IV cosmological study to further constrain the dark energy equation of state \citep{2013PhR...530...87W} - would not be available until 40 such measurements have been carried out with similar precision.

Given these numbers and forecasts, a truly competitive TDC probe would greatly benefit from a much improved accuracy and precision for each lens study. In particular, three sources of uncertainty have been identified as the biggest contributors to the total time-delay distance error budget, i) the time delays, ii) the mass along the line of sight (LOS), and iii) the lens mass parameterisation. Assuming that future time delays can be measured to the percent level - based on long-baseline optical monitoring campaigns and new curve-shifting algorithms \citep{2013A&A...553A.120T} -, any time-delay cosmological probe that aims at obtaining an accuracy and precision of 1-2 percent in the near future, will have to drastically improve their estimate of the external convergence ($\kappa_{\rm ext}$) associated with LOS mass distributions and lift the modelling degeneracy due to different lens mass parameterisations.

Interestingly, \citet{2017ApJ...836..141M} developed a new framework to model LOS mass distributions efficiently and quantified the environmental effects through realistic simulations of lens fields. By reconstructing the three-dimensional mass distribution of strong-lens sightlines, they obtain constraints on $\kappa_{\rm ext}$ that are consistent with those from statistical approaches of combining galaxy number density observations with N-body simulations \citep{2007MNRAS.382..121H,2009A&A...499...31H,2013MNRAS.432..679C,2014ApJ...788L..35S}, but with a 4 times narrower distribution which yields much stronger
priors on $\kappa_{\rm ext}$ (which affects \h\ linearly). Progress in reducing the uncertainty in the lens mass parameterisation, on the other hand, has been moderate. Stellar kinematic data of the foreground lens are now commonly employed to break the MSD and to align the time-delay distance measurements when e.g. a Navarro-Frank-White \citep[NFW;][]{1996ApJ...462..563N,1997ApJ...490..493N} or power law profile are adopted for the lens mass model. But, follow-up kinematic observations from adaptive-optics (AO) assisted ground-based facilities struggle to go beyond a single aperture-averaged velocity dispersion measurement, due to the difficulty in separating the bright quasar light from the foreground lens galaxy and the faintness of the lens itself. As a consequence, the final precision on \h\ is currently limited to $\sim$ 7\% from a single lens system. Moreover, time-delay studies employing kinematic data assume a spherical mass model for recovering the aperture-averaged stellar velocity dispersion, whereas the strong lensing mass model is of elliptical or even triaxial nature, and are thus not fully self-consistent. Whether this assumption introduces a bias in the inferred time-delay distances also remains to be seen.

To drastically improve the precision and accuracy of a single lens study, more flexible dynamical models along with both high-spatially resolved observations that map the 2D stellar kinematics in great detail and sufficient signal-to-noise (S/N) to reliably extract the kinematic moments across the entire field-of-view (FOV) are necessary. The next generation of telescopes - such as the James Webb Space Telescope (\jwst), the Thirty Meter Telescope (\tmt) and the European Extremely Large Telescope (\elt) - will provide the required improvement in sensitivity and resolution. The aim of this paper is to present a fully self-consistent, physically motivated modelling machinery for TDC, that will be capable of exploiting this data set to its full potential. Given the specifications of \jwst, we will create mock stellar kinematics and, based on a joint strong lensing \& stellar dynamical analysis, forecast the cosmological constraints from future space- and ground-based telescope observations.

The paper is organised as follows. In Section \ref{sec:theory}, we cover the strong lensing and stellar dynamical theory and formalism. Section \ref{sec:data} will be used to present the already available \hst\ observations, time delays and mock future stellar kinematics of the prominent strong lens configuration \rxj. We model the data of \rxj\ within a Bayesian framework in Section \ref{sec:analysis}, show the probability density function (PDF) for its time-delay distance and lens distance (\dd) and discuss possible sources of uncertainty. The inference of the cosmological parameters is carried out in Section \ref{sec:results} and finally followed by a summary in Section \ref{sec:summary}.

Throughout this paper, we adopt a standard cosmological model with \h\ $ = 82.5$ \kmsM, a matter density of $\Omega_{\mathrm{m}} = 0.27$ and a dark energy density of $\Omega_{\mathrm{\Lambda}} = 0.73$, where our particular choice for \h\ $ = 82.5$\,\kmsM is driven by the time-delay distance measurements of \rxj\ in \cite{2014ApJ...788L..35S}.

%============================= Section 2 =============================
\section{Theory}
\label{sec:theory}
%=====================================================================

%---------------------------------------------------------------------
\subsection{Historical context}
\label{sec:context}
%---------------------------------------------------------------------

Strong gravitational lensing and stellar dynamics are powerful tracers of their underlying gravitational potential. Being subject to the mass-sheet degeneracy (in lensing) and mass-anisotropy degeneracy (in stellar dynamics), however, it has been quickly realised that a combination of both would be capable of lifting their inherent modelling degeneracies, while providing even tighter constraints for the respective mass models of any given system \citep{1996ApJ...464...92G,1999ApJ...516...18R,2004astro.ph.12596K}. Early implementations of joint strong lensing and stellar dynamical models can also be found in \cite{2002ApJ...568L...5K} and \cite{2002ApJ...575...87T}, where strong lensing and stellar kinematic data have been utilised to infer the internal mass distribution and dark matter content of intermediate-redshift ($z\le1$) galaxies \citep{2003ApJ...583..606K,2004ApJ...611..739T}, culminating in the SLACS survey \citep{2006ApJ...638..703B}. These early studies mainly relied on \hst\ imaging and (aperture averaged) spectroscopic data within the effective radius, while adopting spherical Jeans models. Ultimately, this approach has also been commonly applied for cosmological purposes \citep{2002MNRAS.337L...6T,2003ApJ...599...70K}, to break the mass profile degeneracies in the lensing-only models.

With the advent of integral field spectroscopy and further refinement in lens \citep[e.g.][]{2001ASPC..237...65B,2003ApJ...590..673W,2005MNRAS.363.1136K} and dynamical modelling machineries \citep[e.g.][]{1999ApJS..124..383C,2008MNRAS.390...71C}, the joint analysis has been expanded to cover valuable information from the extended Einstein rings and 2D kinematics, while employing pixelated source reconstruction and more sophisticated two-integral \citep{2007ApJ...666..726B,2009MNRAS.393.1114B,2011MNRAS.415.2215B} and three-integral dynamical models \citep{2010ApJ...719.1481V,2012MNRAS.423.1073B}. Whereas the earliest implementations treated the subject in an inconsistent manner, fully decoupling the strong lensing and stellar kinematic data by, e.g. adopting elliptical and spherical mass models respectively and making simplistic assumptions about the orbital anisotropy profile \citep{1979PAZh....5...77O,1985MNRAS.214P..25M,1985AJ.....90.1027M}, later models have been capable to remedy most of these shortcomings by treating the subject within a fully self-consistent \citep{2010ApJ...719.1481V} and statistically meaningful \citep{2012MNRAS.423.1073B} framework. Here, we build on the work by \cite{2010ApJ...719.1481V} and \cite{2012MNRAS.423.1073B}, by extending the machinery to include time-delay data for cosmological purposes. Moreover, even with the next generation of ground- and space-based stellar kinematic data, the likelihood functions will be swamped by the lensing information, which is why we make use of the Bayesian Information Criterion \citep[BIC;][]{schwarz1978} as a statistical tool to further break degeneracies in future TDC studies.

Our joint strong lensing and stellar dynamical modelling machinery relies on a pixelated source fitting algorithm and the solutions of the Jeans equations in axisymmetric lens geometry, which are embedded in a
Bayesian framework. For brevity, we refer the reader to \cite{2006MNRAS.371..983S,2010ApJ...711..201S,2013ApJ...766...70S} and \cite{2008MNRAS.390...71C}, which cover in detail the theory and application of each to real observational data. Here, we confine ourselves to a description of the main formalisms, following the strong lensing and stellar dynamical framework developed and formulated in \cite{1985A&A...143..413S,1992grle.book.....S,1986ApJ...310..568B} and \cite{1987gady.book.....B}.

%---------------------------------------------------------------------
\subsection{Time-delay strong lensing}
\label{sec:stronglensing}
%---------------------------------------------------------------------

In any strong lens configuration with a time-variable background source, an excess time delay
\begin{equation}
\label{eqn:eqn1}
t(\vec{\theta},\vec{\beta}) = \frac{(1+z_{\rm d})}{c}\;\frac{D_{\rm d} D_{\rm s}}{D_{\rm ds}}\;\phi(\vec{\theta},\vec{\beta})
\end{equation}
will be observed. Here, $\vec{\theta}$ is the angular image position, $\vec{\beta}$ the corresponding source position, $z_{\mathrm{d}}$ the lens redshift, $D_{\rm d}$, $D_{\rm s}$ \& $D_{\rm ds}$ the angular diameter distance to the lens, the source and between the lens and source respectively, and
\begin{equation}
\label{eqn:eqn2}
\phi(\vec{\theta},\vec{\beta}) = \Big[ \frac{(\vec{\theta}\;-\;\vec{\beta})^2}{2}-\psi_{\rm L}({\vec{\theta}}) \Big]
\end{equation}
the Fermat potential. The difference in the light propagation time at image position $\vec{\theta}$ with respect to the non-lensed case can therefore be attributed to the first and second term in the Fermat
potential, which represent the geometric excess path length and the gravitational time-delay of the lens potential $\psi_{\rm L}(\vec{\theta})$ respectively, and a combination of cosmological distances which are generally referred to as the time-delay distance
\begin{equation}
\label{eqn:eqn3}
\dt \equiv (1+\zd)\;\frac{\dd \ds}{\dds}.
\end{equation}
With \dt\ being inversely proportional to \h, Eq.\ref{eqn:eqn1} can be rewritten as
\begin{equation}
\label{eqn:eqn4}
t({\vec{\theta},\vec{\beta}}) \propto \frac{1}{H_0}\;\phi(\vec{\theta},\vec{\beta}),
\end{equation}
i.e., the excess time delay can be used as a cosmological probe, if the form of the lens potential is sufficiently well known. However, since $t(\vec{\theta},\vec{\beta})$ itself cannot be measured, we rely on
relative time delays
\begin{equation}
\label{eqn:eqn5}
\Delta t_{ij} = t_{i} - t_{j} \propto \frac{1}{H_0}\:\Big[\phi(\vec{\theta_{i}},\vec{\beta})\;-\;\phi(\vec{\theta_{j}},\vec{\beta})\Big]
\end{equation}
between multiple images $i$ and $j$, with e.g. quadruply imaged systems naturally providing more constraints than doubly lensed sources.

Given the observables $\vec{\theta_{i}}$ and $\Delta t_{ij}$, the lens potential $\psi_{\mathrm{L}}$ and the source position $\vec{\beta}$ need to be modelled accurately to infer \h. A major drawback of this inference, though, is the MSD\footnote{A special case of the Source-Position Transformation \citep{2014A&A...564A.103S, 2017A&A...601A..77U}.} \citep{1985ApJ...289L...1F}. For illustration purposes, we assume a transformation of the lens potential $\psi_{\mathrm{L}}$ of the form
\begin{equation}
\label{eqn:eqn6}
\psi_{\mathrm{L},\lambda}(\vec{\theta}) = \frac{\lambda}{2}\;|\vec{\theta}|^2+\vec{s}\;\vec{\theta}+c_0+(1-\lambda)\;\psi_{\mathrm{L}}(\vec{\theta}),
\end{equation}
where $\lambda$, $c_0$ and $\vec{s}$ are constant scalars and vectors respectively.  Moreover, the projected matter density $\rho_{\mathrm{2D}}$ is related to its gravitational potential via Poisson's equation
\begin{equation}
\label{eqn:eqn7}
\nabla^{2}\psi_{\mathrm{L}} = 4\pi G\rho_{\mathrm{2D}} = 2\kappa,
\end{equation}
where
\begin{equation}
\label{eqn:eqn8}
\kappa = \frac{\Sigma(D_{\mathrm{d}}\vec{\theta})}{\Sigma_{\mathrm{crit}}}
\end{equation}
is the projected dimensionless surface mass density (SMD) and
\begin{equation}
\label{eqn:eqn9}
\Sigma_{\mathrm{crit}} = \frac{c^2}{4\pi G}\;\frac{\ds}{\dd\dds}
\end{equation}
the critical SMD, that is used to discriminate between the weak ($\kappa \lesssim 1$) and strong lensing regime ($\kappa \gtrsim 1$). According to Eqs. \ref{eqn:eqn6} and \ref{eqn:eqn7}, any transformation $\psi_{\mathrm{L},\lambda}(\vec{\theta})$ of $\psi_{\mathrm{L}}(\vec{\theta})$ will translate into a transformed source position
\begin{equation}
\label{eqn:eqn10}
\beta_{\lambda} = (1-\lambda)\;\beta - \vec{s}
\end{equation}
and a dimensionless SMD of the form
\begin{equation}
\label{eqn:eqn11}
\kappa_{\lambda} = \lambda + (1-\lambda)\;\kappa(\vec{\theta}).
\end{equation}
That is, as $\lambda$, $c_0$ and $\vec{s}$ only change the position and scaling of the source, which by itself is not directly observable, the above transformation essentially implies that any transformation of the lens potential (and hence of the dimensionless lens SMD) will be compensated by a corresponding scaling in the source plane, leaving many observables invariant under the transformation. Unfortunately, however, \dt\ does not belong to this set of invariants. Being highly susceptible to the MSD, \dt\ will also be scaled as follows
\begin{equation}
\label{eqn:eqn12}
D_{\Delta t, \lambda} = \frac{\dtmod}{(1-\lambda)},
\end{equation}
with \dtmod\ being the model time-delay distance (without accounting for the mass-sheet-transformation parameter $\lambda$ in Eq. \ref{eqn:eqn11}), and any cosmological inference based on strong lensing alone is therefore fundamentally limited by our ignorance of $\lambda$.

Since lensing is sensitive to all mass along the LOS, including small and large scale structures in the projected vicinity which can contribute to the SMD at the lens location, the MSD is inherently linked to this external convergence (\kext). In fact, the MSD stems from the degeneracy between \kext\ and the normalisation of the lens potential \citep[but see][for a critical discussion]{2013A&A...559A..37S}. Consequently, Eq. \ref{eqn:eqn12} reduces to
\begin{equation}
\label{eqn:eqn13}
D_{\Delta t} = \frac{\dtmod}{(1-\kext)},
\end{equation}
where \dt\ is the true time-delay distance to the specific sightline of the lens, after accounting for the MSD. In contrast to $\lambda$, though, \kext\ has the benefit of not simply being an arbitrary scaling of the lens potential, but being observationally and/or numerically assessable via photometric and spectroscopic observations of the lens environment \citep{2006ApJ...642...30F,2006ApJ...641..169M,2017MNRAS.467.4220R,2017MNRAS.470.4838S} as well as ray-tracing methods through e.g. the Millennium Simulations
\citep{2007MNRAS.382..121H,2009A&A...499...31H,2013ApJ...768...39G,2013MNRAS.432..679C} and weak lensing \citep{2018MNRAS.477.5657T}. Whereas early attempts to quantify \kext\ have been only moderately successful, yielding external convergences that can affect the final measurement of \h\ by 5\% and more, more recent studies indicate that the distribution and impact of \kext\ can be drastically reduced when e.g. sightlines are not significantly overdense \citep{2013ApJ...768...39G, 2017MNRAS.467.4220R} or individual lens fields are modelled \citep{2017ApJ...836..141M}.

Nonetheless, other means are needed to effectively break the MSD, and to reliably measure \dt. This is particularly evident from Eq. \ref{eqn:eqn4} and \ref{eqn:eqn7}. Assuming, for instance, a
simple power law profile for the 3D density distribution (i.e. $\rho_{\mathrm{3D}}(\vec{r}) \propto r^{-\gamma}$), the lens potential becomes $\psi_{\mathrm{L}} \propto r^{2-\gamma}$. Any uncertainty in the slope of the mass density will thus translate into an uncertainty on the inferred time-delay distance ($\dt \propto \frac{1}{\gamma-1}$).  To constrain $\gamma$, methods have been developed to make use of the spatially extended lensed images of the source galaxy, i.e. the Active Galactic Nuclei (AGN) host galaxy in the case of time-delay lenses \citep[e.g.][]{2003ApJ...590..673W,2006MNRAS.371..983S,2008MNRAS.388..384D,2015ApJ...813..102B}. In this work, we follow the work of \citet{2006MNRAS.371..983S} and \citet{2013ApJ...766...70S} for the lens modelling by describing the AGN host galaxy surface brightness on a grid of  pixels, and the lens mass distribution with parameterised profiles. In addition, stellar kinematics of the lens galaxy are employed, which provide an independent assessment of the lens potential at different radii, further breaking lens mass model degeneracies for constraining \dt. 

%---------------------------------------------------------------------
\subsection{Axisymmetric Jeans modelling}
\label{sec:jeansmodelling}
%---------------------------------------------------------------------

The dynamical state of a system of particles is fully described by its distribution function (DF) $f(\vec{x},\vec{v}) \ge 0$, with particle positions $\vec{x}$ and velocities $\vec{v}$. In the case that these
particles are collisionless, interact purely via gravitational forces and are embedded in background potential $\psi_{\rm D}$ that is smooth in time and space, the time evolution of the DF is subject to the
Collisionless Boltzmann Equation (CBE) \citep{1987gady.book.....B}
\begin{equation}
\label{eqn:eqn14}
\frac{\partial f}{\partial t} + \sum_{i=1}^{3}\;v_i\;\frac{\partial f}{\partial x_i} - \frac{\partial \psi_{\rm D}}{\partial x_i}\;\frac{\partial f}{\partial v_i} = 0,
\end{equation}
which basically postulates a conservation of the phase-space density. Yet, as the phase-space distribution is not accessible for objects beyond our Galaxy, where only bulk motions and positions of stars
along specific LOSs are observed, the CBE is impractical for real observational purposes. In fact, any real world application of the CBE would require a more practical formalism, incorporating kinematic moments which are more easily measurable via line profile shifts and widths. This can be achieved by multiplying the CBE with powers of the velocity moment and subsequent integration over velocity space. Further, rewriting the CBE in terms of the cylindrical coordinate system ($R,z,\phi$) and under the assumption of axial symmetry, we obtain the two Jeans equations
\citep{1922MNRAS..82..122J,1987gady.book.....B}
\begin{equation}
\label{eqn:eqn15}
\frac{\nu\overline{v_{R}^{2}} - \nu\overline{v_{\phi}^{2}}}{R}+\frac{\partial(\nu\overline{v_{R}^{2}})}{\partial R}+\frac{\partial(\nu\overline{v_{R}v_{z}})}{\partial z} = -\nu\;\frac{\partial\psi_{\mathrm{D}}}{\partial R}
\end{equation}
\begin{equation}
\label{eqn:eqn16}
\frac{\nu\overline{v_{R}v_{z}}}{R}+\frac{\partial(\nu\overline{v_{z}^{2}})}{\partial z}+\frac{\partial(\nu\overline{v_{R}v_{z}})}{\partial R} = -\nu\;\frac{\partial\psi_{\mathrm{D}}}{\partial z},
\end{equation}
where $\nu(\vec{x}) = \int fd^{3}\vec{v}$ is the zeroth velocity moment and tracer density of the gravitational potential $\psi_{\mathrm{D}}$\footnote{$\psi_{\mathrm{D}}$ is the 3D gravitational potential, in contrast to $\psi_{\mathrm{L}}$ which denotes the 2D lens potential.}.

Given the four unknown second-order velocity moments $\overline{v_{i}^{2}} = \frac{1}{\nu}\int fv_{i}^{2}d^{3}v_{i}$ and $\overline{v_{i}v_{j}} = \int v_{i}v_{j}fd^{3}v$, the Jeans equations do not have a unique solution. In practice, assumptions about the shape and alignment of the velocity ellipsoid are made to simplify Eq. \ref{eqn:eqn15} and \ref{eqn:eqn16}. These usually include the alignment of the velocity ellipsoid with the cylindrical coordinate system (i.e. $\overline{v_{R}v_{z}} = 0$) and a flattening in the
meridional plane, i.e. $\beta_{z} = 1-{\overline{v_{z}^{2}}}/{\overline{v_{R}^{2}}}$ \citep{1980MNRAS.190..873B,1982MNRAS.200..361B,1987gady.book.....B}, which yield the more commonly seen form \citep{2008MNRAS.390...71C}
\begin{equation}
\label{eqn:eqn17}
\frac{\beta_{z}\nu\overline{v_{z}^{2}}-\nu\overline{v_{\phi}^{2}}}{R}+\frac{\partial(\beta_{z}\nu\overline{v_{z}^2})}{\partial R} = -\nu\;\frac{\partial\psi_{\mathrm{D}}}{\partial R}
\end{equation}
\begin{equation}
\label{eqn:eqn18}
\frac{\partial(\nu\overline{v_{z}^{2}})}{\partial z} = -\nu\;\frac{\partial\psi_{\mathrm{D}}}{\partial z}.
\end{equation}
These equations now link a mass and tracer density to three intrinsic second-order velocity moments which, in turn, can be used to obtain a projected second-order velocity moment along the LOS
\begin{equation}
\label{eqn:eqn19}
\begin{split}
\overline{v_{\mathrm{LOS}}^{2}} = &\frac{1}{\mu(x',y')}\;\int_{-\infty}^{\infty}\;\nu\Big[(\overline{v_{R}^{2}}\sin^{2}\phi+\overline{v_{\phi}^{2}}\cos^{2}\phi)\sin^{2}i \\
& +\overline{v_{z}^{2}}\cos^{2}i-\overline{v_{R}v_{z}}\sin\phi\sin(2i)\Big]dz'\;\equiv\;v^{2}+\sigma^{2}.
\end{split}
\end{equation}

Here, $x'$ and $y'$ are the cartesian coordinates on the plane of the sky, $z'$ the coordinate along the LOS, $i$ the inclination angle, $\mu$ the observed surface brightness (SB) - in contrast to $\nu$, which represents the (deprojected) intrinsic luminosity density - $\cos\phi = x/R$ (where
$x$ and $R^{2} = x^{2}+y^{2}$ denote the intrinsic coordinate axis and cylindrical radius) and $v$ and $\sigma$ the observed mean LOS velocity and velocity dispersion (with $\vrms = \sqrt{v^{2}+\sigma^{2}}$).

The assumptions for the shape and alignment of the velocity ellipsoid have been found to be a good description of the internal dynamical structure of fast-rotating \citep{2007MNRAS.379..401E} early-type galaxies \citep[ETGs;][]{2007MNRAS.379..418C}. Hence, the axisymmetric Jeans equations provide a decent fit to the observed kinematics and are usually in agreement with constraints that are obtained via more sophisticated orbit-based dynamical models. However, the Jeans equations do not make use of the higher order kinematic moments, which contain valuable information regarding the intrinsic shapes of galaxies. This is most prominent for massive, slow-rotating and pressure supported ETGs, where generally worse fits are obtained as
the assumption of axial symmetry also breaks down \citep{2018ApJ...863L..19L}. In principle, the Jeans equations can be extended to triaxial systems, consisting of three equations and six second-order moments, but the set of solutions still contains unphysical DFs ($f < 0$) \citep{2003MNRAS.342.1056V}. We proceed with these caveats in mind and note that the validity of the axisymmetric Jeans equations needs to be evaluated on a case by case basis, especially for the purposes of precision cosmology, where we aim to constrain the mass profile and hence time-delay distances to the percent level.

When constructing axisymmetric Jeans models, the intrinsic luminosity density is obtained by deprojecting the observed SB distribution. For this, we make use of a Multi-Gaussian Expansion (MGE) \citep{1992A&A...253..366M,1994A&A...285..723E,2002MNRAS.333..400C}\footnote{For the dynamical modelling part, we make use of publicly available Python implementations of JAM \citep[\textsc{jampy};][]{2008MNRAS.390...71C} and MGEfit \citep[\textsc{mgefit};][]{2002MNRAS.333..400C}, online available via \url{https://www-astro.physics.ox.ac.uk/~mxc/software/}.}. In brief, the distribution is parameterised by a set of two-dimensional Gaussians, such that the SB can be written as
\begin{equation}
\label{eqn:eqn20}
\mu(x',y') = \sum_{j=1}^{N}\;\mu_{0,j}\;\exp\Big[-\frac{1}{2\sigma'^{2}_{j}}\;(x'^{2}+\frac{y'^{2}}{q'^{2}_{j}})\Big],
\end{equation}
where $\mu_{0}$ is the peak SB, $\sigma'$ the dispersion along the projected major axis and $q'$ the apparent flattening of each Gaussian. In an oblate axisymmetric case, the inclination $i$ is the only free viewing angle required to perform the deprojection. The deprojection is not unique \citep{1987IAUS..127..397R}, unless the galaxy is viewed edge-on, but konus densities which project to zero SB have been found to be of little effect for SB distributions of realistic (elliptical) galaxies \citep{1997MNRAS.287..543V}. If a deprojectable (i.e. $\cos^{2}i < q'^{2}_{\rm min}$, with $q_{\rm min}$ being the axis ratio of the flattest Gaussian in the fit) inclination $i$ has been chosen, the intrinsic luminosity density in cylindrical coordinates reads as
\begin{equation}
\label{eqn:eqn21}
\nu(R,z) = \sum_{j=1}^{N}\frac{q'_{j}\mu_{0,j}}{\sqrt{2\pi}\sigma'_{j} q_{j}}\;\exp\Big[-\frac{1}{2\sigma'^{2}_{j}}(R^{2}+\frac{z^{2}}{q^{2}_{j}})\Big],
\end{equation}
where $\sigma_{j} = \sigma'_{j}$ and $q_{j} =
\frac{\sqrt{q_{j}^{'2}-\cos^{2}i}}{\sin i}$ now denote the intrinsic dispersion and flattening of the Gaussians. Simple (mass-follows-light) models can easily be constructed by linking the tracer density $\nu$ to the mass density $\rho$ via a mass-to-light ratio (M/L) for the individual Gaussians (\ML$_{j}$). The MGE is
particularly handy here, as the gravitational potential can then be obtained by means of a simple, one-dimensional integral
\begin{equation}
\label{eqn:eqn22}
\psi_{{\rm D},j}(R,z) = -\frac{2G\ML\nu}{\sqrt{2\pi}\sigma'_{j}}\;\int_{0}^{1}\;\mathcal{F}_{j}(u)du
\end{equation}
with
\begin{equation}
\label{eqn:eqn23}
\mathcal{F}_{j}(u) = \exp \Big[-\frac{u^{2}}{2\sigma'^{2}_{j}}\Big(R^{2}+\frac{z^2}{\mathcal{Q}^{2}_{j}(u)}\Big)\Big]\frac{1}{\mathcal{Q}_{j}(u)}
\end{equation}
and $\mathcal{Q}^{2}_{j}(u) = 1-(1-q^{2}_{j})\;u^{2}$.

%---------------------------------------------------------------------
\subsection{Joint formalism \& Bayesian Inference}
\label{sec:formalism}
%---------------------------------------------------------------------

We start our joint formalism with the strong lensing part. Given the image positions $\vec{\theta}$ of the AGN and AGN host, a lens potential $\psi_{\rm L}$ will be adopted, which relates the AGN and AGN host source positions to those in the image plane via the lens equation
\begin{equation}
\label{eqn:eqn24}
\vec{\theta} = \vec{\beta}-\vec{\nabla}\psi_{\rm L}(\vec{\theta}) = \vec{\beta} - \vec{\alpha}(\vec{\theta}),
\end{equation}
where $\vec{\alpha}(\vec{\theta)}$  is the scaled deflection angle.
We describe the source intensity distribution on a grid of pixels with values $\bold{s}$ (vector with dimension $N_{\rm s}$, the number of source pixels), which is related to the observed intensity value of the image plane $\bold{d}$ via
\begin{equation}
\label{eqn:eqn25}
\bold{d} = \mathsf{f}\bold{s}+\bold{n}.
\end{equation}
Here, $\bold{d}$ is a vector with length $N_{\rm d}$ (the number of image pixels) and $\bold{n}$ the noise in the data. $\mathsf{f}$ represents the lensing operator (a matrix of dimension $N_{\rm d}\times N_{\rm s}$),
which contains information regarding the lens potential and observational effects - such as telescope point spread function (PSF) - and which is constrained by the extended image positions and intensities as well as the relative time delays between the individual AGN images. In general, lens potentials for which the deflection
angles $\vec{\alpha}(\vec{\theta})$ (and hence the lensing operator $\mathsf{f}$) can be obtained analytically and/or with only moderate numerical effort are adopted, and the goodness-of-fit for a particular model is defined as
\begin{equation}
\label{eqn:eqn26}
\begin{split}
\chi_{\mathrm{L}}^{2}
\propto\;\exp\Bigg[-\frac{1}{2}&\Big[\sum_{i}\frac{({d_{i}}-d_{i,{\rm
      m}})^{2}}{\sigma_{{\rm d},i}^{2}}+\sum_{j}\frac{(a_{j}-a_{j,{\rm
      m}})^{2}}{\sigma_{{\rm a},j}^{2}}\\
&+\sum_{k}\frac{(\Delta t_{k}-\Delta t_{k,{\rm m}})^{2}}{\sigma_{\Delta t,k}^{2}}\Big]\Bigg].
\end{split}
\end{equation}
In the above equation, $d_{i}$ represents the image pixel intensities, $d_{i,{\rm m}}$ the modelled image pixel intensities and $\sigma_{{\rm d},i}$ the corresponding pixel uncertainties. Whereas the first term in
Eq. \ref{eqn:eqn26} represents the fit to the image intensity distribution, the second and third terms account for the $\chi^{2}$ contribution from fitting in particular to the AGN positions ($a_{j}$) and their relative time delays ($\Delta t_{k}$). A best-fitting model is usually quickly obtained by minimising the cost function in Eq. \ref{eqn:eqn26}, when the parameters of interest are few. However, since we are interested in inferring credible confidence intervals for all parameters of interest, we perform an analysis within the framework of Bayesian statistics.

For simplicity, let us assume that the strong lens configuration is well parameterised by a softened power law elliptical mass distribution \citep[SPEMD;][]{1998ApJ...502..531B} with SMD
\begin{equation}
\label{eqn:eqn27}
\kappa = \Big[\frac{\zeta^{2}+\zeta_{\rm c}^{2}}{E^{2}}\Big]^{\eta/2-1},
\end{equation}
with $\zeta^{2} = x'^{2}+y'^{2}/q^{2}$, where $E$ is a normalisation factor, $\eta$ the power law index, $\zeta_{\rm c}$ the core radius and $q$ the observed flattening with the $x'$-axis being aligned with the projected galaxy major axis.  According to Bayes' theorem, the posterior PDF for this lensing-only model - with the set of parameters $\tau_{\mathrm{L}} = \{E, \eta, \zeta_{\rm c}, q, \omega\}$ and data sets
$d_{\mathrm{L}} = \{{d_{i}, a_{j}, \Delta t_{k}}\}$ - is given by
\begin{equation}
\label{eqn:eqn28}
P_{\mathrm{L}}(\tau_{\mathrm{L}}|d_{\mathrm{L}}) \propto \overbrace{P_{\mathrm{L}}(d_{\mathrm{L}}|\tau_{\mathrm{L}})}^\text{likelihood}\;\underbrace{P_{\mathrm{L}}(\tau_{\mathrm{L}})}_\text{prior},
\end{equation}
where the log likelihood corresponds to Eq. \ref{eqn:eqn26} \citep[$\log P_{\rm L} \propto -\chi^2_{\rm L}/2$, after marginalising over the source intensity pixel parameters $\bold{s}$;][]{2010A&A...524A..94S} and $\omega$ comprises a set of remaining variables, such as \dtmod, \kext\ and the position angle (PA) of the projected SPEMD on the plane of the sky
(we measure counter-clockwise from the $x'$-axis to the $y'$-axis). Once (non-)informative priors for $\tau_{\mathrm{L}}$ have been chosen, the marginalised posterior PDF for a parameter of interest can be obtained by integrating the joint (lensing-only) PDF over all nuisance parameters. Similarly, the dynamics-only posterior PDF
\begin{equation}
\label{eqn:eqn29}
P_{\mathrm{D}}(\tau_{\mathrm{D}}|d_{\mathrm{D}}) \propto P_{\mathrm{D}}(d_{\mathrm{D}}|\tau_{\mathrm{D}})\;P_{\mathrm{D}}(\tau_{\mathrm{D}})
\end{equation}
can easily be obtained by means of the dynamics-only likelihood
\begin{equation}
\label{eqn:eqn30}
\chi_{\mathrm{D}}^{2} = \exp\Big[-\frac{1}{2}\sum_{l}\frac{\left(V_{\mathrm{rms},l}-\sqrt{\overline{v_{\mathrm{LOS},l,m}^{2}}}\right)^{2}}{\sigma^{2}_{V_{\mathrm{rms},l}}}\Big].
\end{equation}
Note that $\tau_{\mathrm{L}} \neq \tau_{\mathrm{D}}$. Unlike the lensing data, the dynamics are insensitive to e.g. \kext, while depending explicitly on additional parameters, such as $\beta_{z}$ and
\dd. In the case of a joint lensing \& dynamics model, the joint prior is simply a union of both $\tau_{\mathrm{LD}} = \{\tau_{\mathrm{L}},\tau_{\mathrm{D}}\}$ and the lensing \& dynamics
posterior PDF reads as
\begin{equation}
\label{eqn:eqn31}
\begin{split}
P_{\mathrm{LD}}(\tau_{\mathrm{LD}}|d_{\mathrm{LD}}) \propto\ & P_{\mathrm{LD}}(d_{\mathrm{LD}}|\tau_{\mathrm{LD}})\;P_{\mathrm{LD}}(\tau_{\mathrm{LD}})\\
& = P_{\mathrm{L}}(d_{\mathrm{L}}|\tau_{\mathrm{LD}})\;P_{\mathrm{D}}(d_{\mathrm{D}}|\tau_{\mathrm{LD}})\;P_{\mathrm{LD}}(\tau_{\mathrm{LD}}),
\end{split}
\end{equation}
given the independence of both data sets.

The exploration of the parameter space and adequate sampling of the joint lensing \& dynamics posterior PDF is carried out by means of the affine-invariant ensemble sampler \textsc{emcee} \citep{2013PASP..125..306F}. Starting points are obtained by first carrying out a pre-annealing process, in order to avoid low probability modes of the multi-parameter space, and initialising the walkers of the sampler such that they sample well the prior probability distribution. At each step, i.e. at each parameter combination that is probed by the walkers, the lensing likelihood in Eq. \ref{eqn:eqn26} will be evaluated. Secondary products of this evaluation are the dimensionless SMD (Eq. \ref{eqn:eqn8}) and SB distribution, which are then transformed into a physical mass and luminosity density profile, before being parameterised by a MGE (Eq. \ref{eqn:eqn20} and
\ref{eqn:eqn21}) to allow for a straightforward, analytical calculation of the lens potential according to
Eq. \ref{eqn:eqn22}. Here, the critical SMD in Eq. \ref{eqn:eqn9} has to be expressed in terms of \dt, \dd\ and the lens redshift \zd
\begin{equation}
\label{eqn:eqn32}
\Sigma_{\mathrm{crit}} = \frac{c^{2}}{4\pi G}\;\frac{\dtmod}{(1-\kext)(1+\zd)\dd}\;\frac{1}{\dd}.
\end{equation}
The SMD profile (as obtained from the lensing part) is multiplied with ($1-\kext$), in order to take into account any contribution from the external convergence (see Eq. \ref{eqn:eqn11}). As a
consequence, \kext\ cancels out in the mass density profile that is used in Eq. \ref{eqn:eqn17} and \ref{eqn:eqn18}, and we are thus insensitive to \kext\ when using the kinematic data. Note, however, that the absolute scaling of the lens potential is fixed and the MSD broken when stellar kinematics are included, which provide an independent measurement of the lens potential. Moreover, as \dd\ itself is insensitive to the external convergence along the LOS when inferred from kinematic and lensing observations \citep[][and also noted above]{2015JCAP...11..033J}, we have
\begin{equation}
\label{eqn:eqn33}
\dd = \ddmod.
\end{equation}

After deprojection and adoption of an anisotropy parameter $\beta_z$, the likelihood function in Eq. \ref{eqn:eqn30} can finally be evaluated. In combination with $\tau_{\mathrm{LD}}$, this yields the joint lensing \& dynamics posterior PDF in Eq. \ref{eqn:eqn31}, where the marginalised distributions can then be visualised by histograms with the most probable model and the 1$\sigma$ uncertainties being approximated by the median and 16th and 84th percentiles of the distribution.

%---------------------------------------------------------------------
\subsection{Bayesian Information Criterion}
\label{sec:formalism}
%---------------------------------------------------------------------

A large, flexible and physically motivated set of light and mass parameterisations is generally utilised to accurately model the foreground lens. This approach is largely motivated by our ignorance of the true underlying gravitational potential, which is tightly linked to the excess time-delay (Eq. \ref{eqn:eqn5}). Even if lens galaxies are found to be well approximated by power law density distributions \citep{2006ApJ...649..599K}, the inferred time-delay distances can differ significantly when compared to density distributions that follow more closely e.g. a NFW profile. The discrepancy in \dt\ between different lens mass models can be partially alleviated by including stellar kinematic information. However, the vast amount of data from both lensing and future IFU stellar kinematics should enable us to perform a model selection, by means of the differences in their likelihood functions. To this end we will make use of the BIC \citep{schwarz1978}, which is an approximation to the Bayesian evidence\footnote{In practice, the BIC approximation performs well, despite an asymptotic error of \textit{O}(1). In cases where the data is of lower quality (i.e. with little constraining power) the impact of the error can still be mitigated by choosing appropriate priors, such that it becomes \textit{O}($n^{-1/2}$), where $n$ is the number of data points \citep{kass1995reference}.} \citep[see e.g.][for a thorough derivation]{raftery1995bayesian} via
\begin{equation}
\begin{split}
\label{eqn:eqn34}
%P_{\mathrm{LD}}(d_{\mathrm{LD}}|\tau_{\mathrm{LD}}) \approx \exp(-\mathrm{BIC}/2).
P_{\mathrm{LD}}(d_{\mathrm{LD}}|\mathcal{M}) =
& \int P_{\mathrm{LD}}(d_{\mathrm{LD}}|\mathcal{M},\tau_\mathrm{LD})\;P_{LD}(\tau_\mathrm{LD}|\mathcal{M})  \;d\tau_\mathrm{LD}\\
& \approx \exp(-\mathrm{BIC}/2),
\end{split}
\end{equation}
where $\mathcal{M}$ is a hypothesis (in our case a model with lensing and dynamical parameters $\tau_{\mathrm{LD}}$).
Following Bayes' theorem, the ratio of posterior probabilities of two competing hypothesis $\mathcal{M}_{1}$ and $\mathcal{M}_{2}$ are given by
\begin{equation}
\label{eqn:eqn35}
\underbrace{\frac{P_{\mathrm{LD}}(\mathcal{M}_{1}|d_{\mathrm{LD}})}{P_{\mathrm{LD}}(\mathcal{M}_{2}|d_{\mathrm{LD}})}}_\text{posterior odds} = \overbrace{\frac{P_{\mathrm{LD}}(d_{\mathrm{LD}}|\mathcal{M}_{1})}{P_{\mathrm{LD}}(d_{\mathrm{LD}}|\mathcal{M}_{2})}}^\text{Bayes factor}\;\underbrace{\frac{P_{\mathrm{LD}}(\mathcal{M}_{1})}{P_{\mathrm{LD}}(\mathcal{M}_{2})}}_\text{prior odds}
%\underbrace{\frac{P_{\mathrm{LD}}(\tau_{\mathrm{LD,1}}|d_{\mathrm{LD}})}{P_{\mathrm{LD}}(\tau_{\mathrm{LD,2}}|d_{\mathrm{LD}})}}_\text{posterior odds} = \overbrace{\frac{P_{\mathrm{LD}}(d_{\mathrm{LD}}|\tau_{\mathrm{LD,1}})}{P_{\mathrm{LD}}(d_{\mathrm{LD}}|\tau_{\mathrm{LD,2}})}}^\text{Bayes factor}\;\underbrace{\frac{P_{\mathrm{LD}}(\tau_{\mathrm{LD,1}})}{P_{\mathrm{LD}}(\tau_{\mathrm{LD,2}})}}_\text{prior odds}
\end{equation}
Accordingly, an approximation to the Bayes factor can be obtained by means of Eq. \ref{eqn:eqn34} and \ref{eqn:eqn35}. Here, in particular, we make use of the BIC differences ($\Delta\mathrm{BIC} = \mathrm{BIC_{1}} - \mathrm{BIC_{2}}$) between model $\mathcal{M}_{1}$ and $\mathcal{M}_{2}$ with model parameters $\tau_{\mathrm{LD,1}}$ and $\tau_{\mathrm{LD,2}}$, which yields
\begin{equation}
\label{eqn:eqn36}
\frac{P_{\mathrm{LD}}(d_{\mathrm{LD}}|\mathcal{M}_{1})}{P_{\mathrm{LD}}(d_{\mathrm{LD}}|\mathcal{M}_{2})} = \exp(-\Delta\mathrm{BIC}/2).
\end{equation}
That is, the approximation to the Bayes factor can readily be utilised to assess the posterior odds, if the prior odds are assumed to be equally probable. Given a larger set of models $\mathcal{M}_{j}$ with BIC values $\mathrm{BIC}_{j}$, the posterior probability for a particular model $i \in j$ follows as
\begin{equation}
\label{eqn:eqn37}
P_{\mathrm{LD}}(\mathcal{M}_{i}|d_{\mathrm{LD}}) = \frac{\exp(-\mathrm{BIC}_{i}/2)}{\sum\limits_{1}^{j} \exp(-\mathrm{BIC}_{j}/2)},
\end{equation}
which in turn can be used as a weighting scheme for model averaging purposes, since according to the law of total probability, $\sum\limits_{1}^{j} P_{\mathrm{LD}}(\mathcal{M}_{j}|d_{\mathrm{LD}}) = 1$.\\

The BIC is given by
\begin{equation}
\label{eqn:eqn38}
\mathrm{BIC} = \mathrm{ln}(n)k - 2\mathrm{ln}(\hat{L}),
\end{equation}
In the above equation, $k$ represents the number of model parameters, $n$ the number of data points and $\hat{L}$ the corresponding model maximum likelihood. The BIC discerns between candidate models by penalising models of increased complexity (i.e. with higher degrees freedom). The model maximum likelihood is given by $P_{\mathrm{L}}(d_{\mathrm{L}}|\tau_{\mathrm{L}})$, $P_{\mathrm{D}}(d_{\mathrm{D}}|\tau_{\mathrm{D}})$ or $P_{\mathrm{LD}}(d_{\mathrm{LD}}|\tau_{\mathrm{LD}})$, depending on which data set is fitted. Yet, owing to the MSD and the minuscule
differences in the likelihood function of different mass parameterisations when modelling \rxj\ \citep{2014ApJ...788L..35S}, the lensing-only likelihood cannot be used to discriminate between models. Even in a joint model, the likelihood would easily be dominated by the large number of pixel intensities that are fitted in the lensing part, which is why we will use the differences in the goodness of fit of the IFU kinematics from our joint strong lensing \& dynamics run, to perform a proper weighting of our models according to the BIC. We follow the approach of \cite{2019MNRAS.484.4726B}, where a weighting scheme with respect to the minimal BIC is defined as
\begin{equation}
\label{eqn:eqn39}
f_{\mathrm{BIC}}(x) = \\
\begin{cases}
1 & x \le \mathrm{BIC_{min}}\\
\exp(-\frac{x-\mathrm{BIC_{min}}}{2}) & x > \mathrm{BIC_{min}}.
\end{cases}
\end{equation}
This weighting scheme follows Eq. \ref{eqn:eqn37}, after accounting for the fact that the calculation of the denominator is cumbersome in most realistic scenarios. Especially for modelling time-delay lenses, where we rely on a relatively small subset of models, which are physically motivated by e.g. hydrodynamical numerical simulations \citep{1996ApJ...462..563N,1997ApJ...490..493N}. Given its purpose of being a normalising factor, we approximate the weighting via Eq. \ref{eqn:eqn39}, while making sure that $f_{\mathrm{BIC}}$ is still bound by 1.

For a given model including the lens mass/light distribution, PSF, AGN light, and AGN host galaxy surface brightness, the BIC value of this model could be computed to rank it relative to other models.  We are particularly interested in comparing the lens mass parametrisation, and thus the changes in BIC due to different mass parameterisation.  However, the BIC depends on also other modelling choices/parameters, especially the number of surface brightness pixels used to describe the AGN host galaxy, which introduces an uncertainty on the BIC \citep[see][where source pixelisation dominates the uncertainties in the BIC for a given form of lens mass parametrisation]{2013ApJ...766...70S}.  Given finite computing resources and thus a finite number of source intensity grids that we could explore, we quantify the uncertainty on the BIC due to the source grid pixelisation effect by comparing the BIC values of a range of source grids and estimating the scatter $\sigma_{\rm BIC}$.  To account for the uncertainty in the BIC in weighting models, we follow \citet{2019MNRAS.484.4726B} and convolve $f_{\mathrm{BIC}}$ in Eq. \ref{eqn:eqn39} with a Gaussian of variance $\sigma_{\rm BIC}^2$, obtaining
the new weights as
\begin{equation}
\label{eqn:eqn40}
f_{\mathrm{BIC}}^{*}(x) = h(x,\sigma_{\rm BIC}) \ast
f_{\mathrm{BIC}}(x),\\
\end{equation}
where
\begin{equation}
\label{eqn:eqn41}
h(x,\sigma_{\rm BIC}) = \frac{1}{\sqrt{2\pi}\sigma_{\rm BIC}} \exp\left(
   -\frac{x^2}{2\sigma_{\rm BIC}^2}\right).
\end{equation}
Carrying out the convolution integral, we find an analytic expression as follows
\begin{equation}
\label{eqn:eqn42}
\begin{split}
f_{\mathrm{BIC}}^{*}(x) = &\frac{1}{2} + \frac{1}{2} {\rm
  Erf}\left( \frac{{\rm BIC}_{\rm min}-x}{\sqrt{2} \sigma_{\rm
      BIC}}\right) \\
& + \frac{1}{2}\exp \left( \frac{1}{2}{\rm BIC}_{\rm min}+\frac{1}{8}\sigma_{\rm
  BIC}^2-\frac{1}{2}x \right) \cdot \\
&\ \ \ \ {\rm Erfc} \left( \frac{2 {\rm BIC}_{\rm
      min}+\sigma_{\rm BIC}^2-2x}{2\sqrt{2} \sigma_{\rm BIC}}\right),
\end{split}
\end{equation}
where the Erf and Erfc functions are defined as
\begin{equation}
\label{eqn:eqn43}
{\rm Erf}(z) \equiv \frac{2}{\sqrt{\pi}}\int_{0}^{z} \exp\left(-t^2\right) {\rm d}t,
\end{equation}
and
\begin{equation}
\label{eqn:eqn44}
{\rm Erfc}(z) \equiv 1- {\rm Erf}(z).
\end{equation}

%============================= Section 3 =============================
\section{Data}
\label{sec:data}
%=====================================================================

In this section, we present real and mock observations of \rxj, which are used to infer future cosmological constraints by means of our joint strong lensing \& stellar dynamical analysis. We focus on \rxj,
in particular, given that it i) is the brightest lens galaxy among the \holi\ base sample, ii) has the most precise time-delay measurements, with only 1.3\% uncertainty in the longest time delay, and iii) has
plenty of ancillary data, which make it the most promising candidate for future integral field unit (IFU) observations.

Our data consist of \hst\ imaging, precise time delay measurements and mock \jwst\ stellar kinematics. Rather than mocking up \jwst\ imaging along with \jwst\ IFU data, we choose to rely on the literature \hst\ data instead. Even if \jwst\ will be able to acquire comparably high S/N observations with much shorter exposure times, \jwst\ NIR imaging will provide only a marginal improvement over \hst, given its slightly smaller nominal pixel size of 0.03\arcsec/pixel and expected PSF FWHM of $\sim$ 2 pixels. More importantly, both \hst\ and \jwst\ are capable of obtaining spatially-resolved 
% observations
imaging
 with sufficient S/N for bright lenses, as in the case of \rxj, to reach the required precision of 0.02 for the total mass density power law slope, such that the uncertainty on the Fermat potential is already subdominant with respect to other sources of uncertainty \citep{2015JCAP...09..059M}. Mocking up ground-based AO imaging from future 30-40m class telescopes would have been a viable alternative for assessing the overall improvements from both, next generation imaging and spectroscopy. However, the move from \hst\ to ground-based AO observations introduces other sources of uncertainty, such as the recovery of the complex PSF; given that current state-of-the-art analysis of AO images from 8-10m class telescopes yield distance constraints comparable to those from \hst\ \citep{2019arXiv190702533C}, we adopt \hst\ imaging for our current study.

%, and does not yield significant gains in precision, even with state-of-the-art 8-10m class telescopes \citep{2019arXiv190702533C}.

%---------------------------------------------------------------------
\subsection{Imaging}
\label{sec:imaging}
%---------------------------------------------------------------------

\hst\ Advanced Camera for Surveys (ACS) data of \rxj\ have been obtained as part of programme GO:9744 (PI:Kochanek). The data set comprises imaging in the F814W and F555W filters, where five exposures
each have been taken with a total integration time of 1980s. For the analysis, we give preference to the F814W imaging, given the fact that the stellar mass-to-light ratios become a weaker function of the underlying stellar populations \citep{2001ApJ...550..212B,2001MNRAS.326..255C}. Moreover, the F814W data shows a clearer separation between the AGN and the spatially extended Einstein ring, whereas the F555W filter is more difficult to model due to diffraction spikes extending into the lensed arcs (but see \cite{2016JCAP...08..020B} for a joint modelling of both bands).

The reduction and combination of the imaging data is performed via the standard \multi\ pipeline \citep{2002PASP..114..144F}, with charge transfer inefficiencies properly taken into account by empirically
tracing back the charge-coupled device (CCD) detector readout mechanism and thus the initial charge distribution \citep{2010PASP..122.1035A}. The images are sky subtracted, corrected for geometric and photometric distortions and cosmic ray cleaned, before drizzled onto a final science frame with 0.05\arcsec/pixel resolution. Flux uncertainties for each pixel are obtained by adding in quadrature Poisson noise from the source and background noise from the sky and detector readout.

\begin{figure}
\begin{center}
\includegraphics[width=0.95\linewidth]{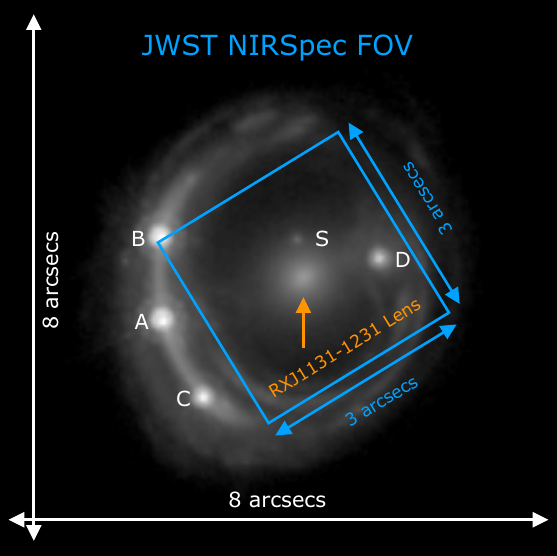}
\end{center}
\caption{\hst\ ACS F814W imaging cutout of \rxj, illustrating the prominent lens configuration with a quadruply imaged background quasar (A, B, C \& D) and a nearby satellite (S). Spectroscopic measurements locate the lens and quasar at redshift $z_{\mathrm{d}} = 0.295$ and $z_{\mathrm{s}} = 0.658$, respectively. Overlaid is the \jwst\ NIRSpec nominal FOV of 3\arcsec$\times$3\arcsec, within which we create mock stellar kinematics of the foreground lens at 0.1\arcsec/pixel resolution. The FOV is oriented such that the x-axis is aligned with the galaxy major axis.}
\label{fig:fig1}
\end{figure}

The final science frame is displayed in Fig. \ref{fig:fig1}, where the centrally located galaxy lenses the background AGN into a quadruple lens configuration (A, B, C \& D). The background quasar host is a spiral
galaxy \citep{2006A&A...451..865C}, which forms the extended Einstein ring. Discovered by \cite{2003A&A...406L..43S}, spectroscopic measurements of the foreground lens and background source yield a redshift of $z_{\mathrm{d}} = 0.295$ and $z_{\mathrm{s}} = 0.654$ \citep{2003A&A...406L..43S,2007A&A...468..885S}, respectively. The foreground lens is further accompanied by a satellite galaxy (S), which is assumed to be a dwarf elliptical \citep{2006A&A...451..865C}.

%---------------------------------------------------------------------
\subsection{Time delays}
\label{sec:time-delays}
%---------------------------------------------------------------------

Time-delay measurements of \rxj\ have been carried out by means of a dedicated optical monitoring campaign within the COSMOGRAIL program. Based on high-cadence (3 days) long-baseline (9+ years and more than 700 epochs) observations with meter-class telescopes, and new curve-shifting techniques, \cite{2013A&A...553A.120T} report time-delays relative to image B of $\Delta t_{\rm AB} = 0.7\pm1.2$ days, $\Delta t_{\rm CB} = -0.4\pm1.5$ days and $\Delta t_{\rm DB} = 91.4\pm1.2$ days, with systematic errors already taken into account in the
uncertainty estimates. In general, the long-baseline measurements result in time delays with $\sim$ 3\% precision \citep{2013A&A...556A..22T,2015ApJ...800...11L,2016A&A...585A..88B}, and microlensing shifts in the time-delays \citep{2018MNRAS.473...80T} are found to be negligible, given the long time delay \citep{2018MNRAS.481.1115C}.
%as they average out over many years.

Given the above measurements, \rxj\ is particularly suitable for TDC purposes. The background AGN is not only quadruply imaged, providing three independent time-delay constraints, but the longest time delay yields an uncertainty of only 1.3\% and forms a comparably low floor for any time-delay distance measurement. While percent level precision of $\Delta t$ is a necessary condition for any TDC probe that aims to measure \h\ to the percent level in a single lens study, it is not sufficient. Even though the time-delay of \rxj\ is the smallest contributor to the error budget in \dt\ \citep{2013ApJ...766...70S}, the final precision of $\sim$ 7\% is still substantial and mostly dominated by uncertainties in the LOS mass contribution and degeneracies in the lens mass model.

%---------------------------------------------------------------------
\subsection{IFU stellar kinematics}
\label{sec:stellarkinematics}
%---------------------------------------------------------------------

Spectroscopic observations of \rxj\ with state-of-the-art instruments have, so far, only been able to yield a single stellar velocity dispersion measurement \citep{2013ApJ...766...70S}, due to the faintness of the lens and difficulties in separating the bright quasar light from the galaxy. Yet, future observatories, such as \jwst\ and \elt, will be capable of obtaining far more than a single aperture averaged measurement of the stellar kinematics, due to their improved sensitivity and resolution. In order to assess the improvements in
constraining \dt, when IFU data from \jwst\ become available, we have mocked up stellar kinematics based on the specifications of \jwst's near-infrared spectrograph (NIRSpec). To this end, we have carried out a lensing-only fit to the imaging data and time-delays of \rxj, with a source resolution of $64\times64$ pixels. The light model consists of four pseudo-isothermal elliptical profiles (PIEMDs), which are used to mimic a two-component \sersic\ distribution \citep{2011MNRAS.417.1621D, 2014ApJ...788L..35S}. The mass model consists of a baryonic and non-baryonic component, where the former is obtained by multiplying the light profile with a constant stellar M/L and the latter is accounted for by a NFW halo. The best-fitting model of this fit is then employed to create a luminosity and mass density profile, and complemented by a minimal set of random dynamical parameters, to create a mock kinematic map of the second-order velocity moment according to Eq. \ref{eqn:eqn19}. Whereas the mock time-delay distance \dtmod\ is based on the best-fitting lensing-only model, the mock input lens distance \ddmod\ is obtained by adopting our standard cosmological model in Sec. \ref{sec:introduction} with $z_{\mathrm{d}} = 0.295$ and $z_{\mathrm{s}} = 0.654$.

\begin{figure}
\begin{center}
\includegraphics[width=0.99\linewidth]{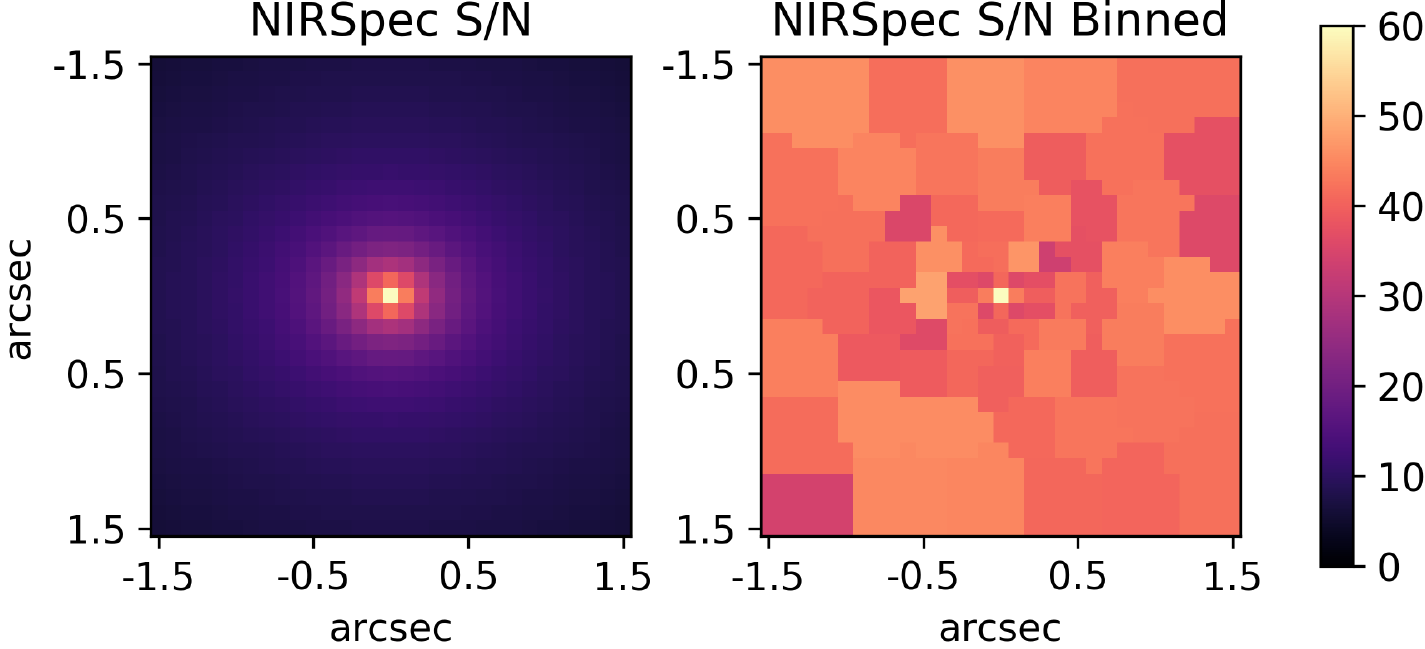}
\end{center}
\caption{Left: SB distribution of \rxj\ at \jwst\ NIRSpec resolution, which has been transformed into a S/N map. The on-source integration time with NIRSpec has been tuned to achieve a S/N of 60 in the central spaxel ($\sim$ 7h with the ETC V1.3). Given the relative intensity distribution from a parameterised fit to its SB profile, the S/N for all spaxels follows accordingly. Right: binned NIRSpec map of \rxj\ to a target S/N of 40 in each bin, to allow for a reliable measurement of the stellar velocity moments across the entire FOV. For simplification, we omitted the AGN images and satellite while mocking up the observations.}
\label{fig:fig2}
\end{figure}

To simulate observational effects and to assess maps of different quality, we spatially bin the map beforehand via an adaptive spatial binning procedure \citep{2003MNRAS.342..345C} to a target S/N of 20, 40 and 60 respectively. Especially the latter two are deemed more than sufficient to reliably extract the kinematics across the entire FOV \citep{2017A&A...597A..48F}. The binning is achieved by assuming a S/N in the central (brightest) pixel. Given the parameterised SB distribution, and assuming a Poisson noise dominated regime, the relative pixel intensities can then be translated into a relative 2D S/N map, which is then binned according to the above requirement (Fig. \ref{fig:fig2}). The S/N in the central pixel is obtained by means of \jwst's exposure time calculator (ETC V1.3), where we aimed for a S/N that is both high enough to yield enough spatially resolved measurements and achievable with reasonable on-source integration times. Our final data sets are comprised of three different S/N combinations. A central S/N of 60 (or 100 or 30, respectively), with a target S/N of 40 (or 60 or 20, respectively) in each bin. The combination of the central and target S/N levels are denoted as 60/40 (100/60 \& 30/20) hereafter, implying a binning scheme where the central S/N is e.g. chosen to be 60 with a target S/N of 40 across all bins.

The mock kinematics cover a 3\arcsec$\times$3\arcsec\ FOV for \jwst\ at 0.1\arcsec/pixel resolution (see Table \ref{table:table1}). This mimics a single pointing with JWST, centred on the lens, where a small cycle dither pattern with subarcsecond shifts can be carried out to allow for identification and removal of cosmic rays and detector defects. The FOV contains $\sim$ 900 spaxels. In reality, however, the number of useful spaxels (and hence the final number of bins) will be smaller than the total number within the nominal FOV, due to
contamination from B, D and S. These will be masked during the fitting of the stellar spectra and extraction of the stellar kinematics. The loss in spatial information, though, should be minimal given the small
point spread function (PSF). Nonetheless, we will also probe a smaller FOV of 2\arcsec$\times$2\arcsec, to quantify the changes in our cosmological constraints when less data is available. This smaller FOV is a conservative assumption and results in a considerable loss of spatial information when compared to the nominal FOV of 3\arcsec$\times$3\arcsec, but can be considered as a worst case scenario, where we aim to predict the minimal gain in our cosmological inference. Note that these mock maps are created to harness the full power of \jwst's IFU spectrograph and are in contrast to previous studies of spatially resolved, but only radially averaged profiles of the velocity dispersion \citep{2018MNRAS.473..210S}.

\begin{table}
	\caption{Specifications of \jwst's IFU mode for mock IFU stellar kinematics of \rxj }
	\begin{center}
	\centerline{
	\begin{tabular}{ c  c }
		\hline
		\hline
		 & \jwst \\
		\hline
		Instrument & NIRSpec \\
		Pixel size & $0.1\arcsec\times0.1\arcsec$ \\
		Field-of-view & $3\arcsec\times3\arcsec$ \\
		PSF FWHM & 0.08\arcsec \\
		Filter & G140H/F100LP \\
		Resolving power & $\sim$ 2700 \\
                \hline
	\end{tabular}}
\end{center}
Notes. The PSF size is the actual size we have used for mocking and modelling purposes, and roughly $2\times$ larger than the diffraction limited PSF FWHM. The mocked up observations can be obtained with the respective filter combination, which is selected such that it covers our target Ca II triplet stellar absorption features, given \rxj's redshift of $z=0.295$. The high resolution grating with its resolving power corresponds to an instrumental velocity dispersion of $\sim$ 50 \kms, likely sufficient to yield reliable measurements of the LOS velocity distribution across the entire FOV.
%	}
%	\vspace{2ex}
	\label{table:table1}
%	\end{center}
\end{table}

\begin{figure*}
\begin{center}
\includegraphics[width=0.99\linewidth]{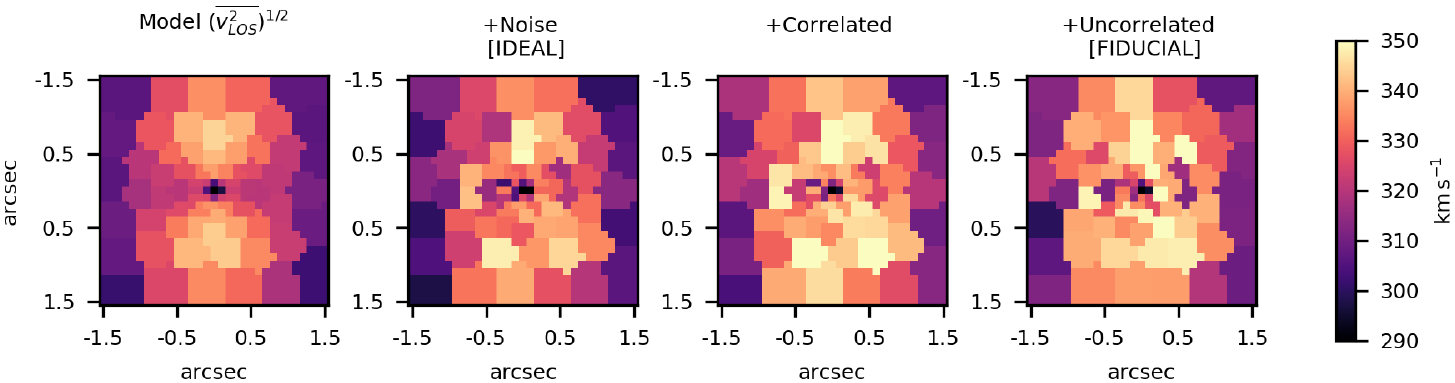}
\end{center}
\caption{Mock IFU \vlos\ maps of \rxj, at \jwst\ NIRSpec resolution, for a S/N configuration of 60/40. First: mock kinematics without any errors. The data is based on the best-fitting lensing-only model, complemented by a minimal set of random dynamical parameters. Second: mock kinematics with random Gaussian errors. The error in each bin depends on its S/N and mock $(\overline{v_{\mathrm{LOS}}^{2}})^{1/2}$ value. This constitutes the ``IDEAL'' data set, without systematic uncertainties in the kinematic measurements (due to e.g., stellar template mismatch).  Third: correlated errors have been added to the noisy $(\overline{v_{\mathrm{LOS}}^{2}})^{1/2}$ map in the second column, which results in a systematic floor of 2\%. Fourth: uncorrelated errors of 2\% have been adopted and added to the map in the third column. This last column illustrates our final mock kinematics, which account for various sources of (systematic) uncertainties and observational difficulties and which will be used throughout our analysis as our reference data set, labelled ``FIDUCIAL''.}
\label{fig:fig3}
\end{figure*}

The kinematic data is convolved with a single Gaussian PSF of size 2$\times$ the diffraction limit of \jwst\ and has a FWHM of 0.08\arcsec. In reality, more complex shapes, that follow more closely e.g. a Moffat profile, can be expected. But, we are not worried about the actual shape of the real \jwst\ observations, as any shape can also easily be adopted by similarly expanding the PSF with a set of multiple Gaussians, to cover e.g. the extended PSF wings. We add realistic errors to the mock kinematics, where the error in each bin is derived by drawing a random number from a Gaussian distribution with $\mu = 0$ and $\sigma_{\rm stat} = (\overline{v_{\mathrm{LOS},l}^{2}})^{1/2}\times\frac{1}{\textrm{(S/N)}_{l}}$ (where ($\overline{v_{\mathrm{LOS},l}^{2}})^{1/2}$ is the mock $(\overline{v_{\mathrm{LOS}}^{2}})^{1/2}$ value at bin position $l$ and $\textrm{(S/N)}_{l}$ its corresponding S/N)\footnote{The signal and noise in an IFU spaxel is commonly defined as the median flux and flux standard deviation across the spectral range. Here, we make the assumption that the S/N scales inversely proportional with the errors of our kinematic measurements \citep{2011MNRAS.414..888E}.}. This standard deviation is employed as our \textit{true} measurement error. Moreover, correlated and uncorrelated uncertainties of 2\% each have been added on top of the random Gaussian noise, to account for observational errors that can arise due to e.g. stellar template mismatches. The former simulates a systematic floor in our mock data set, where we utilise the median $(\overline{v_{\mathrm{LOS}}^{2}})^{1/2} \times\ \frac{1}{50}$ across all bins to offset all measurements by a constant value; the latter follows the approach described above by adding again random Gaussian noise with $\mu = 0$ and $\sigma_{\rm uncorr} =
(\overline{v_{\mathrm{LOS,l}}^{2}})^{1/2}\times\frac{1}{50}$ to all bins across the entire FOV. In summary, we have for each bin
\begin{equation}
\label{eqn:eqn42}
(\overline{v_{\mathrm{LOS,data}}^2})^{1/2}=(\overline{v_{\mathrm{LOS}}^2})^{1/2}+\delta v_{\rm stat}+\delta v_{\rm corr} + \delta v_{\rm uncorr}, 
\end{equation}
where $\delta v_{\rm stat}=$ Gaussian[0,$\sigma_{\rm stat}$], $\delta v_{\rm corr}= 0.02\,(\overline{v_{\mathrm{LOS}}^2})^{1/2}$, and $\delta v_{\rm uncorr} =$ Gaussian[0,$\sigma_{\rm uncorr}$].
The IFU kinematic maps are shown in Fig. \ref{fig:fig3}, where the last column depicts our final mock data, which includes all sources of uncertainty and is employed as our reference data set throughout our joint analysis. 

%============================= Section 4 =============================
\section{Analysis}
\label{sec:analysis}
%=====================================================================

We construct time-delay strong lensing and stellar dynamical models within the axisymmetric Jeans formalism, as described in Sec. \ref{sec:theory}, and make use of the high-resolution \hst\ data, time-delays and mock IFU stellar kinematic maps in Sec. \ref{sec:data}.

%---------------------------------------------------------------------
\subsection{Setup}
\label{sec:setup}
%---------------------------------------------------------------------

To reliably constrain cosmological distances to the percent level, we require flexible and accurate prescriptions of the underlying lens mass distribution. In our joint modelling, we therefore make use of
observationally and theoretically motivated mass models. In the case of \rxj, this includes a \texttt{SPEMD} (which accounts for both the dark and luminous mass) and a \texttt{COMPOSITE} mass model \citep{2014ApJ...788L..35S}, consisting of parameterised fits to the baryonic and non-baryonic matter distribution. Here, the baryonic mass is parameterised via four PIEMDs, which are used to mimic a two-component \sersic\ contribution, and a NFW dark halo, and is therefore identical in nature with the mass model that has been used to mock up the IFU kinematics in Sec. \ref{sec:stellarkinematics}. Both mass models have been shown to provide excellent fits to the strong lensing data \citep{2014ApJ...788L..35S}, but an aperture averaged stellar velocity dispersion measurement was essential to bring both \dt\ distributions into agreement. While the discrepancy could be resolved by including stellar kinematic data, the precision and accuracy is still limited, and this mass model degeneracy is the main contributor to the error budget, which we aim to constrain further by modelling the mock IFU data set.

For simplicity, we neglect the satellite when mocking up the IFU map (Sec. \ref{sec:data}) as well as during the modelling of the strong lensing and stellar kinematic data. The satellite is small enough to result in a loss of only a few spaxels, when being masked during the extraction of the IFU kinematics. More importantly, though, the satellite has a negligible effect for our mass model and cosmological inference \citep{2013ApJ...766...70S}, as it contributes as little as 1\% to the SMD at the lens location.
Our analysis relies on two separate mass models, as described above. The modelling parameters for both mass
parameterisations are presented in Table \ref{table:table2}, along with the dynamical modelling parameters, which have also been used to mock up the kinematic data set. Note that the PIEMDs (i.e. the \sersic\ profiles) are fixed during the fitting process. Fits to the SB distribution are carried out beforehand, and
the SB distribution is translated into a SMD profile by means of a variable stellar M/L. Moreover, the PA and centroids of the dark and luminous matter distribution in the \texttt{COMPOSITE} mass model are fixed to the same value, to ensure that the projected SMD can be deprojected to an intrinsically axisymmetric mass distribution.

\begin{table*}
	\caption{Model parameters and priors for our joint strong
          lensing \& dynamical models, including the cosmological
          distances, the \texttt{SPEMD}, the \texttt{COMPOSITE} mass distribution and
          the dynamical variables. The mock IFU data set is based on
          the best-fitting \texttt{COMPOSITE} lensing-only model with a source
          resolution of $64\times 64$ pixels and random values for the
          dynamical parameters. The mock cosmological distances are
          based on the best-fitting lensing-only model for \dt\ and
          assuming \h\ $ = 82.5$ \kmsM, $\Omega_{\mathrm{m}} = 0.27$,
          $\Omega_{\mathrm{\Lambda}} = 0.73$, $z_{\mathrm{d}} = 0.295$ and $z_{\mathrm{s}} = 0.654$
          for determining \dd.} 
	\begin{center}
	\centerline{
	\begin{tabular}{ c  c  c  c  c }
		\hline
		\hline
		Description & Parameters & Mock input values & Prior type & Prior range\\
		\hline
		\textbf{Distances} & & & & \\
		Model time-delay distance [Mpc] & \dtmod\ & 1823.42 & Flat & [1000, 3000]\\
		Model lens distance [Mpc] & \ddmod\ & 775.00 & Flat & [600, 1000]\\
		\hline
		\textbf{SPEMD} & & & & \\
		Flattening & $q$ & & Flat & [0.2, 1.0]\\
		Einstein radius [arcsec] & $\theta_{\mathrm{E}}$ & & Flat & [0.01,2.0] \\
		Power law slope & $\gamma'$ & & Flat & [0.2, 0.8]\\
		External shear strength & $\gamma_{\mathrm{{ext}}}$ & & Flat & [0.0, 0.2]\\
		External shear position angle [\textdegree] & $\phi_{\mathrm{ext}}$ & & Flat & [0.0, 360.0]\\
		\hline
		\textbf{COMPOSITE} & & & & \\
		Stellar M/L [\Msun/\Lsun] & \MLstar & 2.09& Flat & [0.5, 2.5]\\
		Flattening & $q$ & 0.73 & Flat & [0.2, 1.0]\\
		Einstein radius [arcsec] & $\theta_{\mathrm{E}}$ & 0.20 & Flat & [0.01,2.0] \\
		Scale radius [arcsec] & $r_{\mathrm{s}}$ & 22.53 & Gaussian & [18.6, 2.6]\\
		External shear strength & $\gamma_{\,\mathrm{ext}}$ & 0.08 & Flat & [0.0, 0.2]\\
		External shear position angle [\textdegree] & $\phi_{\mathrm{ext}}$ & 1.42 & Flat & [0.0, 360.0]\\
		\hline
		\textbf{Dynamics} & & \\
		Anisotropy & $\beta_z$ & 0.15 & Flat & [$-0.3$, 0.3]\\
		Inclination [\textdegree] & $i$ & 84.26 & Flat & [80.0, 90.0]\\
	\end{tabular}
	}
%	\vspace{2ex}
	\label{table:table2}
	\end{center}
\end{table*}

Our respective models include a total of 9 variable parameters with mostly uniform priors for both the \texttt{SPEMD} and \texttt{COMPOSITE} mass models. These priors are weakly informative in a sense that they are bound to ranges around the mock input value, where we have used observationally motivated minimum and maximum values for e.g. the anisotropy $\beta_z$ \citep[e.g.][]{2007MNRAS.379..418C} and power law slope $\gamma'$ \citep[e.g.][]{2010ApJ...724..511A}. The prior ranges are sufficiently large to explore well the PDF, while still allowing for a fast convergence. Note, though, that flat priors are not always suitable for all parameters in the fit, as e.g. the intrinsic shape distribution (and consequently the inclination under the assumption of axisymmetry) is well known to be described by a Gaussian profile \citep[e.g.][]{2014MNRAS.444.3340W}. However, the inclination of \rxj\ is severely limited by the high flattening of the second \sersic\ in the fit to the SB distribution and, as a consequence, we have employed a flat prior for the possible range of deprojections. Moreover, the choice of our particular priors has been tested extensively and found to be insignificant for the inference presented in Sec. \ref{sec:modelling}. Whereas this indicates that the data is indeed powerful enough to draw credible conclusions from the posterior distribution, irrespective of the prior choice, it cannot be considered as evidence for "truly" non-informative priors in the sense of e.g. Jeffreys priors. We therefore advise to probe the impact of such "naive" non-informative prior assumptions, as the posterior can be highly susceptible to the prior choice in less constraining cases.\\
While the \texttt{COMPOSITE} model has an additional variable parameter ($r_{\rm s}$), this parameter is not constrained at all, given that it lies well beyond the coverage of the lensing and kinematic data. Using a Gaussian prior instead \citep{GavazziEtal07}, the two mass models (\texttt{SPEMD} and \texttt{COMPOSITE}) have effectively the same number of free parameters. This number is significantly smaller than the total number of variables in the final lens model of \rxj\ in \cite{2013ApJ...766...70S}. However, most of those variables have a negligible effect on the key cosmological parameters, such as \dt\ and \dd, and we therefore adopt those optimised variables as fixed values during the modelling process. The only exception with respect to the optimised values in \cite{2014ApJ...788L..35S} is the PA of the NFW halo and the \texttt{SPEMD}, which are now both aligned with the SB distribution. As axial symmetry is a necessary condition for the construction of the Jeans models, a vastly different PA would violate our underlying modelling assumption. Source resolutions of varying sizes, on the other hand, will be probed, given that the parameter constraints show significant shifts depending on the pixelisation scheme for the AGN host galaxy surface brightness. Besides the uncertainty due to different mass parameterisations, this systematic uncertainty is, in fact, one of the biggest sources of uncertainty for a given mass model, resulting in a distribution that can be $2 - 3 \times$ as wide as for a fixed source grid resolution \citep{2013ApJ...766...70S}. For each mass model, we will therefore examine eight different source grid resolutions of $54\times54$, $56\times56$, $58\times58$, $60\times60$, $62\times62$, $64\times64$, $66\times66$ and $68\times68$ pixels. These source grid resolutions are usually sufficient to achieve an adequate $\chi^{2}$, while stabilising the final modelling constraints towards a common value. The source grid size is chosen such that it contains the entire source intensity distribution, with the outermost source grid pixels converging towards zero intensity values. Our final lensing-only models will then equally weight the constraints from models of different source grid resolutions and different lens mass parameterisations (i.e. \texttt{COMPOSITE} and \texttt{SPEMD}), by combining the individual Markov Chains. An equal weighting is applied in the lensing-only case, since we are incapable of differentiating between different models due to the MSD. In the case of a joint lensing \& dynamics run, however, a weighting scheme according to the BIC will be applied, where the dynamical likelihood $P_{D}(d_{D}|\tau_{D})$ will be utilised to perform a model selection.

%---------------------------------------------------------------------
\subsection{Modelling}
\label{sec:modelling}
%---------------------------------------------------------------------

\begin{figure}
\begin{center}
\includegraphics[width=0.99\linewidth]{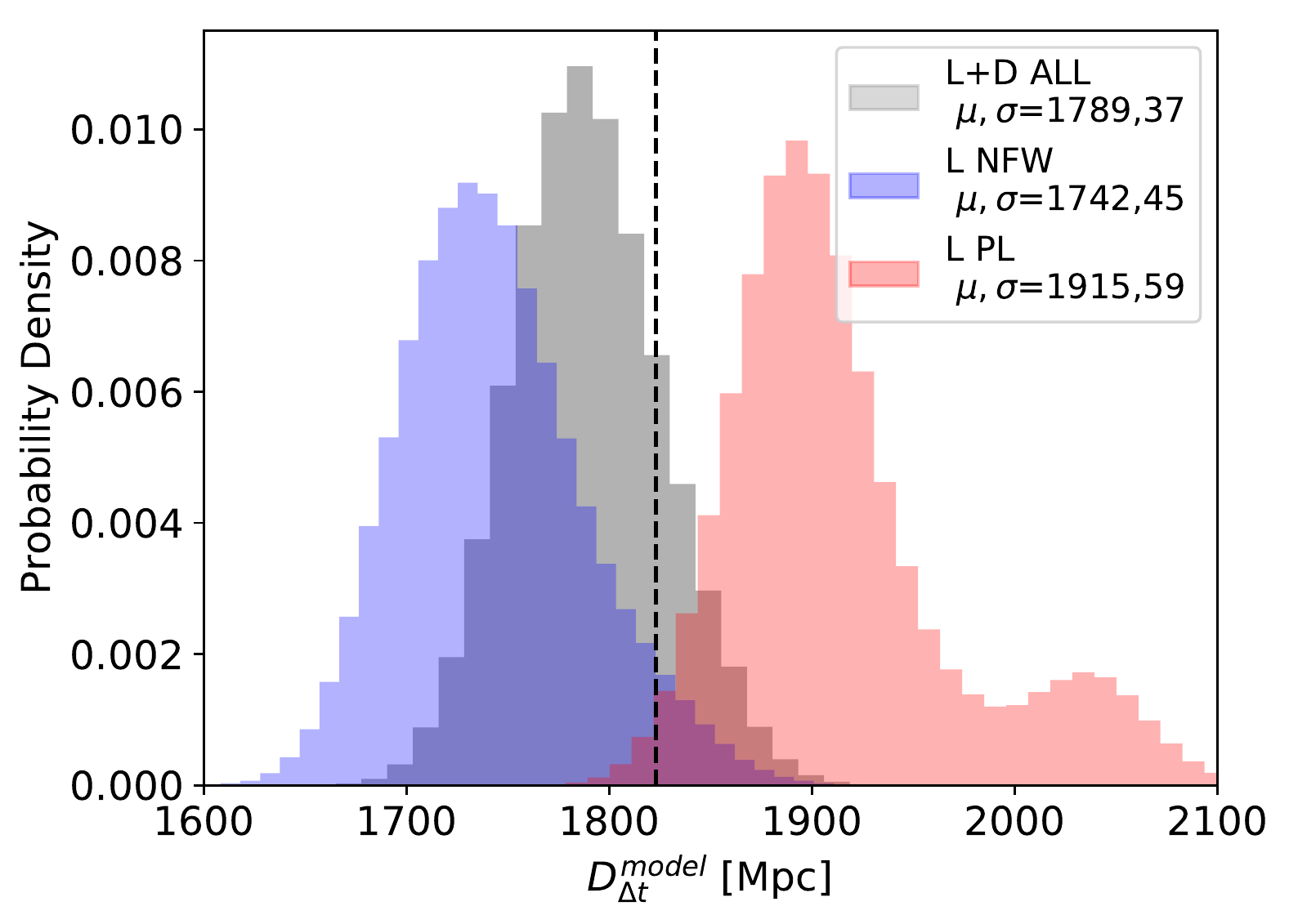}
\includegraphics[width=0.99\linewidth]{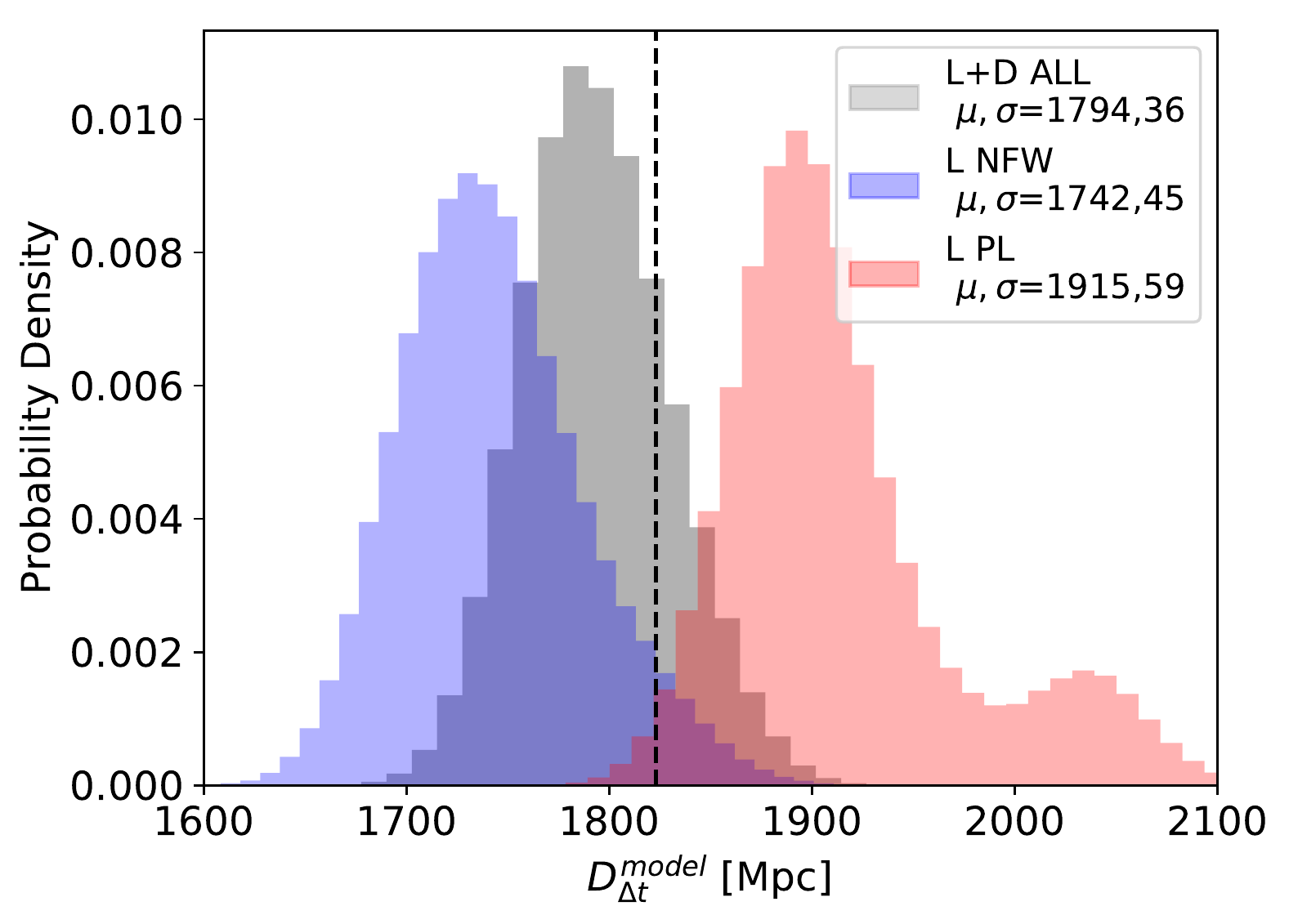}
\end{center}
\caption{Top: marginalised PDF for \dtmod, based on joint strong lensing and stellar dynamical models, for the IDEAL data with statistical noise only and a S/N of 60/40. The blue shaded region shows the PDF for strong lensing-only models and a \texttt{COMPOSITE} mass distribution, including six different source resolutions with equal weighting. The  red shaded region shows the corresponding PDF for strong lensing-only models with a \texttt{SPEMD}. The grey shaded region shows the combined and BIC weighted constraints from the joint run, including both mass parameterisations and all source resolutions. Bottom: same as above, but for our FIDUCIAL kinematic data set, including all sources of uncertainty. The vertical dashed line denotes the best-fitting lensing-only value for the \texttt{COMPOSITE} mass model and a source resolution of $64\times64$ pixels, which was used as input to mock up the IFU kinematics.}
\label{fig:fig4}
\end{figure}

\begin{figure}
\begin{center}
\includegraphics[width=0.99\linewidth]{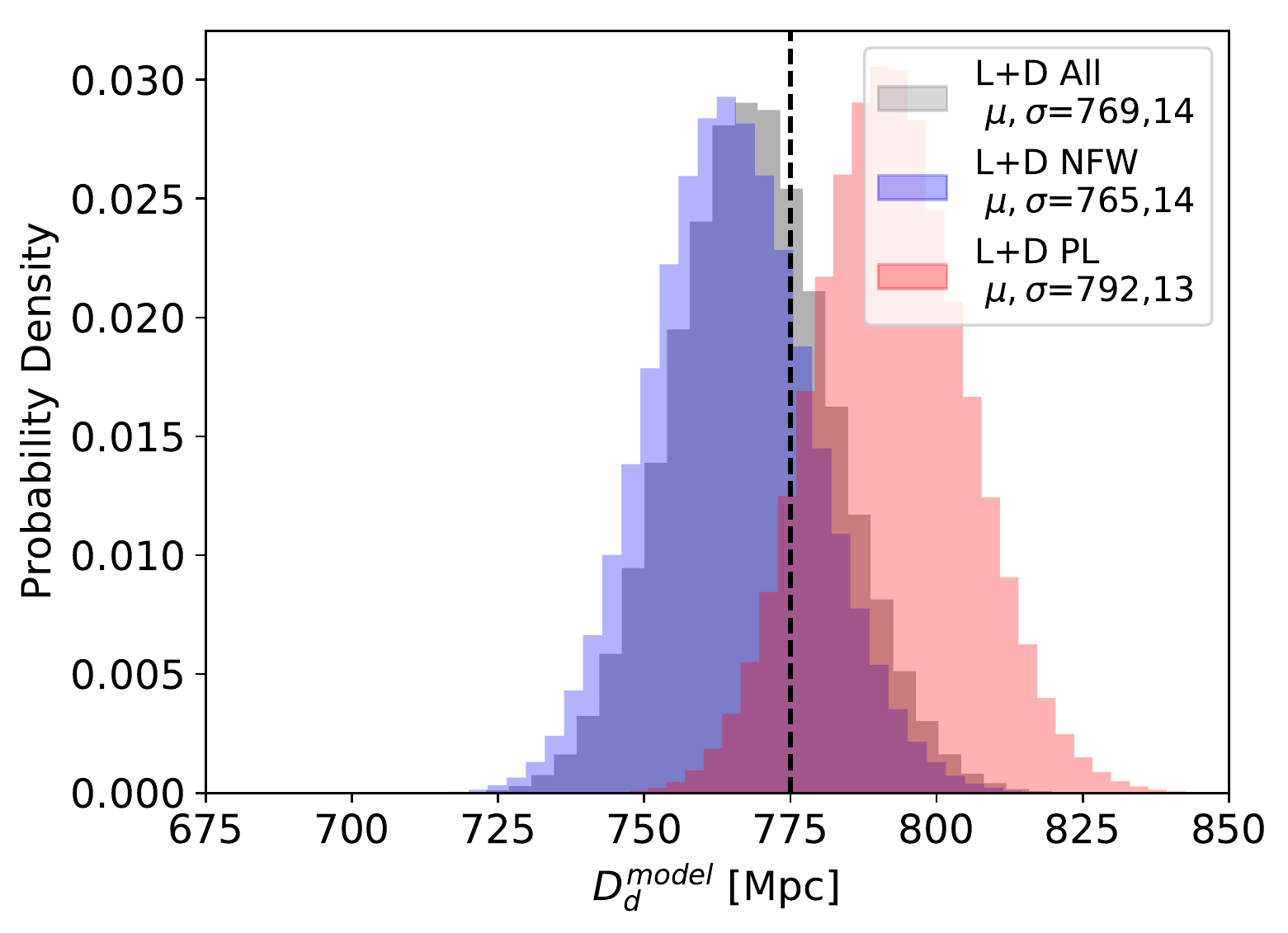}
\includegraphics[width=0.99\linewidth]{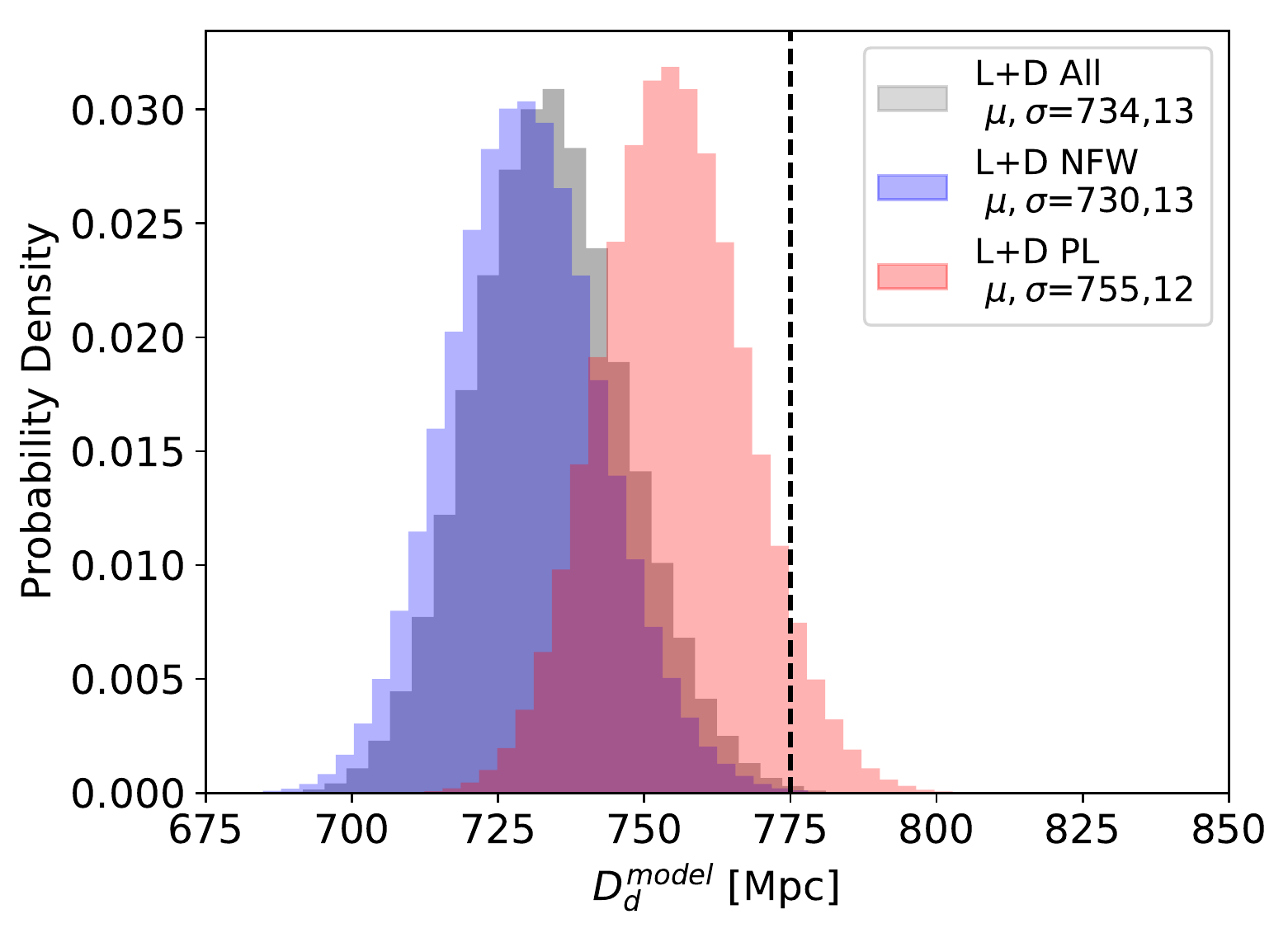}
\end{center}
\caption{Top: marginalised PDF for \ddmod, based on joint strong lensing and stellar dynamical models, for the IDEAL data with statistical noise only and a S/N of 60/40. The blue shaded region shows the PDF for models with a \texttt{COMPOSITE} mass distribution, including six different source resolutions with equal weighting. The red shaded  region shows the corresponding PDF for models with a \texttt{SPEMD}. The grey shaded region shows the combined and BIC weighted constraints from the joint run, including both mass parameterisations and all source resolutions. Bottom: same as above, but for our FIDUCIAL kinematic data set, including all sources of uncertainty. The vertical dashed line denotes the mock lens distance as obtained from the \texttt{COMPOSITE} mass model, a source resolution of $64\times64$ pixels and under the assumption of our chosen cosmology. All models include the stellar kinematic data, as the lensing-only models are insensitive to \ddmod\ alone.}
\label{fig:fig5}
\end{figure}

We visualise the results of our joint strong lensing \& stellar dynamical models in Fig. \ref{fig:fig4} and \ref{fig:fig5}, where we show the marginalised 1D PDFs for our main parameters of interest, i.e. \dtmod\ and \ddmod\, with a S/N of 60/40. The top panels display the constraints from fits to the strong lensing and kinematic data with statistical noise only (hereafter IDEAL, corresponding to the second panel of Fig.~\ref{fig:fig3}), whereas the bottom panels show the results from fits to the IFU stellar kinematics including various sources of uncertainty (hereafter FIDUCIAL, corresponding to the fourth panel of Fig.~\ref{fig:fig3}).
The blue shaded region displays the PDF for models with a \texttt{COMPOSITE} mass distribution, the red shaded region for models with a \texttt{SPEMD} and the grey shaded region the combined PDF from both distributions when the IFU stellar kinematics are included in the fit and a weighting according to the BIC is performed. Note that the blue and red shaded region in Fig. \ref{fig:fig4} is the PDF from lensing-only models. This is in contrast to Fig. \ref{fig:fig5}, where these also include the stellar kinematics, as the lensing-only models are insensitive to \ddmod.

\begin{figure}
\begin{center}
\includegraphics[width=0.99\linewidth]{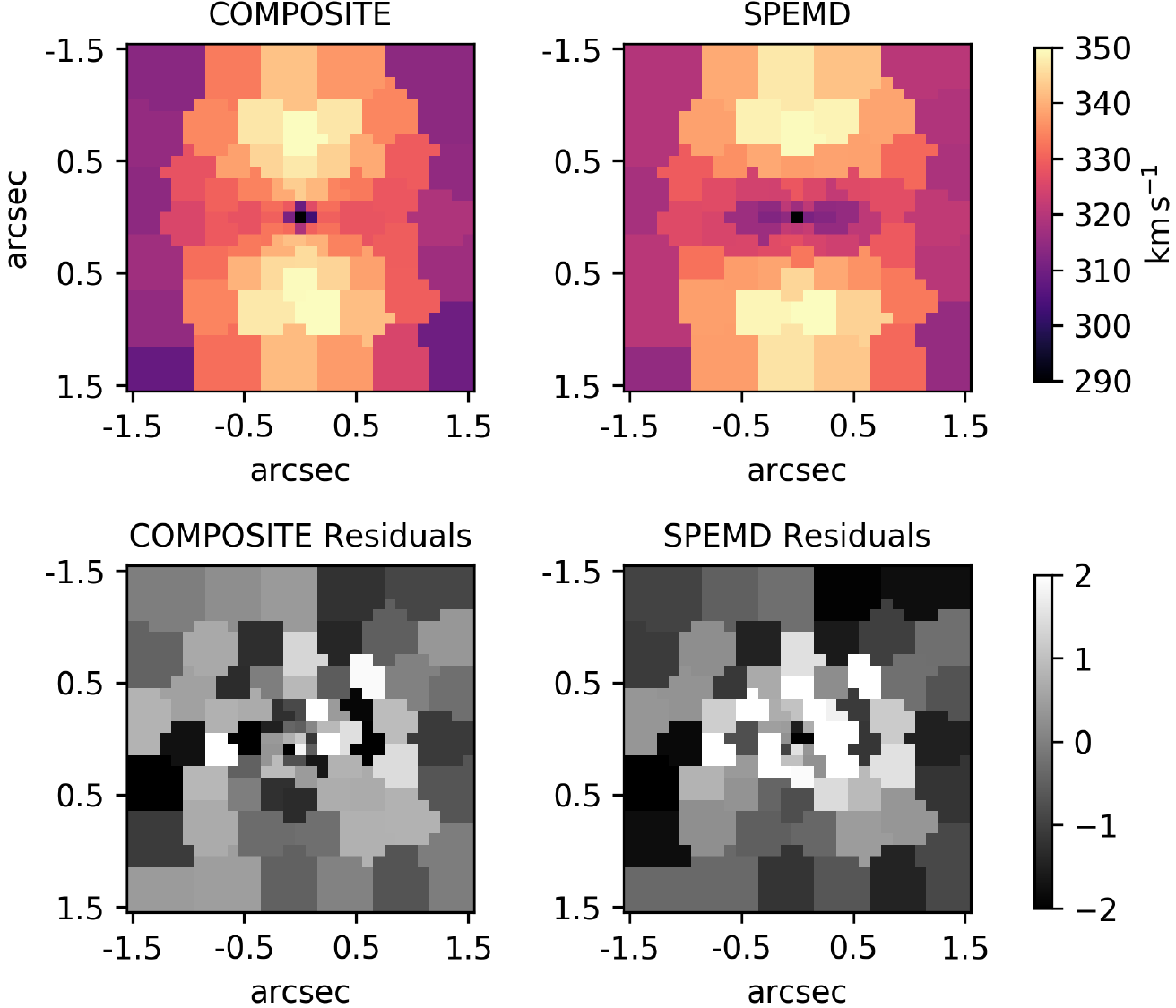}
\end{center}
\caption{Top: mock IFU stellar kinematics of \rxj\ at \jwst\ NIRSpec resolution. The panels show the predicted \vlos\ for the best-fitting joint \texttt{COMPOSITE} and \texttt{SPEMD} model respectively, when fitting to the FIDUCIAL data. Bottom: Residual map for the best-fitting joint \texttt{COMPOSITE} and \texttt{SPEMD} model, normalised by the errors, showing the goodness of fit on a bin by bin basis. The systematic \vlos\ offset in the \texttt{SPEMD} models is used to perform an effective model selection according to the BIC, and results in a significant downweighting of its corresponding probabilities.}
\label{fig:fig6}
\end{figure}

Both panels in Fig.~\ref{fig:fig4} clearly show the discrepancy in the time-delay distance, when different mass parameterisations are employed to model the foreground lens, with a double peaked distribution. Since these two models yield a comparable goodness of fit (a manifestation of the MSD), both are equally plausible and an equal weighting is applied in the lensing-only case. The combined \texttt{COMPOSITE} and \texttt{SPEMD} PDF therefore results in a distribution of $1835^{+89}_{-116}$\,Mpc (here, the 50th and 16th/84th percentiles of the distribution rather than the mean and standard deviation of a single Gaussian). The discrepancy in the time-delay distance measurement can be resolved by including IFU stellar kinematics, where the combined and weighted distributions converge towards the common mock input value of 1823\,Mpc. The 2D kinematics clearly distinguish between both mass models, where the systematic offset in \vlos\ is utilised to significantly downweight the contribution from the \texttt{SPEMD} (Fig.~\ref{fig:fig6}). 
%Note that the stellar kinematics are not capable of aligning both (the \texttt{COMPOSITE} and \texttt{SPEMD}) \dtmod\ distributions, which could be due to the limited freedom in the radial density profile of the power law model \citep{2018MNRAS.474.4648S}.
A Gaussian fit to the combined and weighted \texttt{COMPOSITE} and \texttt{SPEMD} PDF with stellar kinematics included (i.e. grey shaded region) yields [$\mu_{\rm I},\sigma_{\rm I}$] = [1789\,Mpc, 37\,Mpc] in the IDEAL case and [$\mu_{\rm F},\sigma_{\rm F}$] = [1794\,Mpc, 36\,Mpc] for our FIDUCIAL data set respectively. This time-delay distance constraint is a considerable improvement in precision (2.0\%) when compared to the lensing-only models (5.6\%),
even when systematic uncertainties in the stellar kinematics are generously taken into account.

\begin{figure}
\begin{center}
\includegraphics[width=0.99\linewidth]{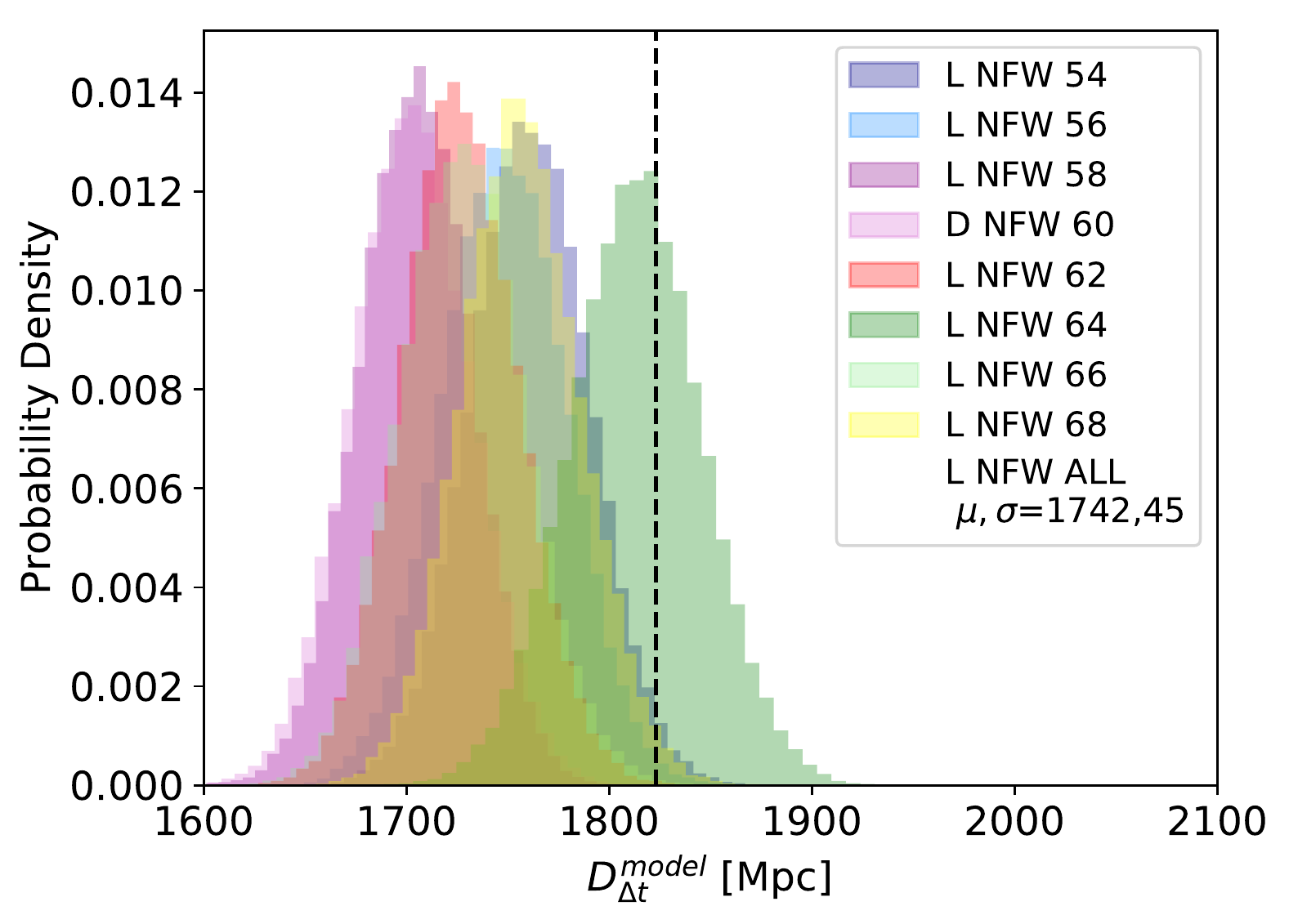}
\includegraphics[width=0.99\linewidth]{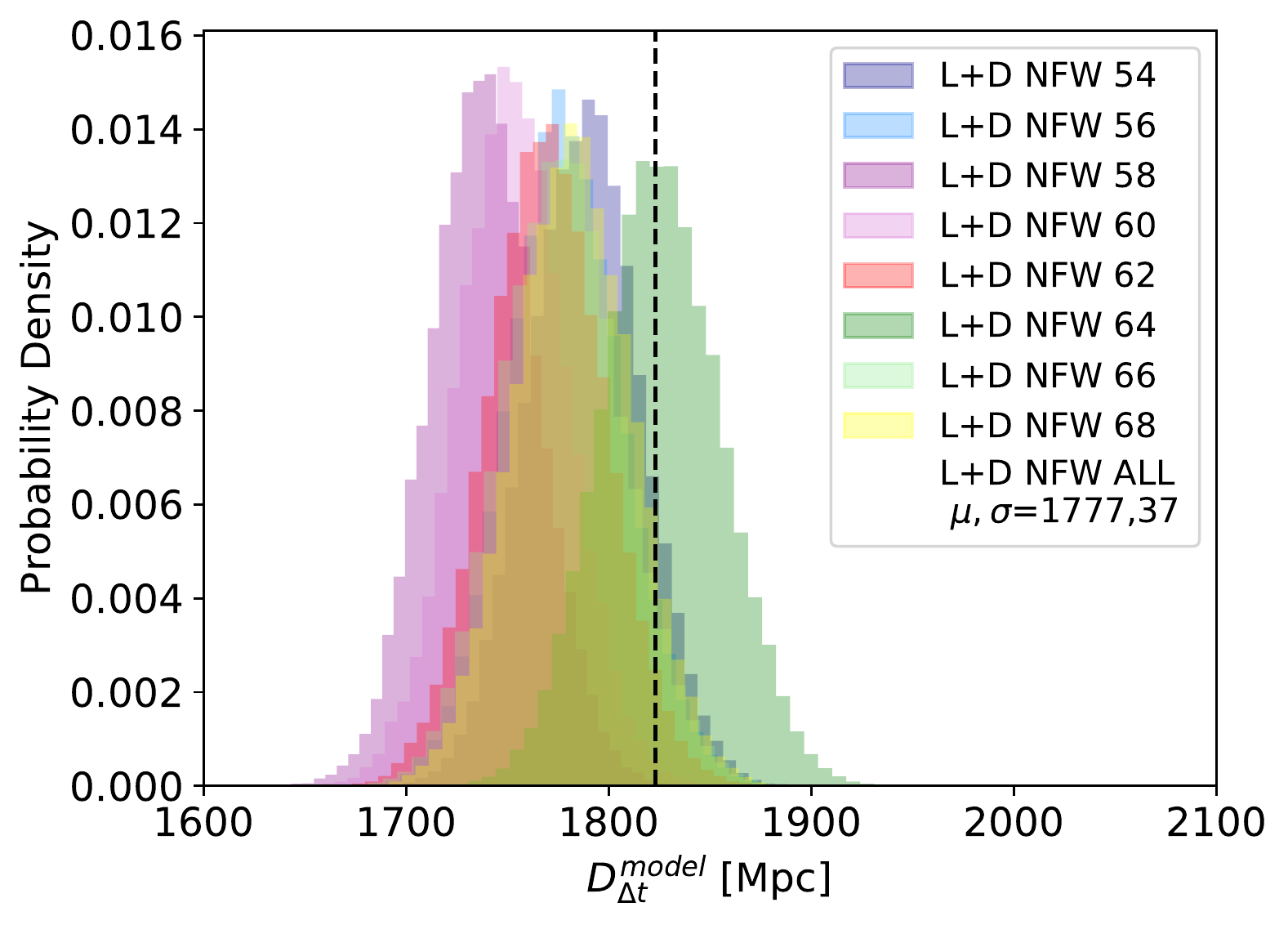}
\end{center}
\caption{Top: marginalised PDFs for \dtmod, based on strong lensing-only models. The individual colours represent the PDFs for models with different source resolutions and the same \texttt{COMPOSITE} mass parameterisation for the foreground lens mass distribution. The mean and standard deviation of the combined PDFs (i.e. for all source resolutions and equal weighting) is given in the legend. Bottom: same as above, but for joint strong lensing and stellar dynamical models of the FIDUCIAL data, where systematic errors in the measurement of the stellar kinematics are included. Information from IFU data helps in reducing the width of the PDF for a given mass parameterisation, getting rid of the low probability wings, and shifts the mean towards the mock input value (vertical dashed line) at a source resolution of $64\times64$ pixels.}
\label{fig:fig7}
\end{figure}

The improvement can be traced back to three effects, in particular, i) a smaller width of the PDF for individual mass parameterisations with different source resolutions, ii) a shift of the mean of the
distribution towards the \textit{true} input time-delay distance and iii) a drastic downweighting of models with a significantly worse goodness of fit. In Fig \ref{fig:fig7}, we illustrate the first two effects by showing the PDFs from lensing-only and joint strong lensing and stellar dynamical models. In both cases, we adopted a \texttt{COMPOSITE} mass distribution and modelled with six different source resolutions. The joint fit to the IFU stellar kinematics considerably reduces the width of the combined PDF from different source resolutions, effectively erasing the low probability wings from lensing-only models in Fig. \ref{fig:fig4}, while shifting the whole distribution towards the input time-delay distance.

A joint fit with stellar kinematics of even higher S/N almost perfectly recovers the input time-delay distance while reaching a precision of 1.7\%. In contrast, lower S/N kinematics yield not only worse precision (of 2.5\%) but also worse accuracy, where the time-delay distance is only recovered within 1.2$\sigma$. The differences in the
constraints for models of different S/N can be attributed to the aforementioned three effects, which are less prominent when the quality of the IFU kinematics degrades.

When it comes to the constraints for the lens distance \ddmod, we observe a much tighter distribution for our reference S/N of 60/40. Fits to the data with statistical noise only recover remarkably well the input lens distance of 775\,Mpc. A Gaussian fit to the combined and weighted PDF from models with a \texttt{COMPOSITE} and \texttt{SPEMD}
yields [$\mu_{\rm I},\sigma_{\rm I}$] = [769\,Mpc, 14\,Mpc]. This is a 1.8\% precision measurement for the secondary cosmological distance we aim to infer. Yet, the distribution is clearly biased towards lower
distances for fits to our FIDUCIAL data set with [$\mu_{\rm F},\sigma_{\rm F}$] = [734\,Mpc, 13\,Mpc], which is a consequence of the correlated noise and systematic floor we have added to $(\overline{v_{\mathrm{LOS}}^{2}})^{1/2}$ to mock realistic observational errors. The bias is especially prominent for models with a \texttt{COMPOSITE} mass distribution, where the joint PDF across all source resolutions (with equal weighting) yields a distribution with [$\mu,\sigma$] = [730\,Mpc, 13\,Mpc]. Keep in mind, however, that the mock data has been created with a source resolution of $64\times64$ pixels, which is at the edge of the joint \texttt{COMPOSITE} PDF for both \dt\ and \dd\ (see e.g. Figure \ref{fig:fig7}). Picking a mock source resolution, and hence an input lens distance, which is closer to the median of the joint PDF in the first place would have partially alleviated this strong bias for our \texttt{COMPOSITE} mass models. Despite our mock source resolution, we observe a similar bias in the joint \texttt{COMPOSITE} and \texttt{SPEMD} PDF across all S/N. Nonetheless, we
are able to recover the \textit{true} lens distance from our joint \texttt{COMPOSITE} and \texttt{SPEMD} PDF within $\sim 3\sigma$, even for our highest S/N pick (with correlated and uncorrelated systematics included),
where the model uncertainties are the tightest.
We also observe that the \dt\ constraints are overall preserved, irrespective of the various kinematic noise properties and realisations. This is attributable to the fact that \dt\ is mainly anchored by the lensing data and time delay measurements. In contrast, information regarding \dd\ is directly embedded in our mock stellar kinematics. As a consequence, it is highly sensitive to any systematic changes in the data. In fact, $(\overline{v_{\mathrm{LOS}}^{2}})^{1/2} \propto 1/(\dd)^{1/2}$. The median $(\overline{v_{\mathrm{LOS}}^{2}})^{1/2}$ difference between the IDEAL and FIDUCIAL data across all bins amounts to $\sim$ 2.1\%, which perfectly explains the $\sim$ 4\% shift in the mean \dd\ between both distributions. We note that the inference of $H_0$ from the joint constraint of $\dt$ and $\dd$ depends on the kinematic measurements, given the dependence of $\dd$ on kinematics (as illustrated in Section \ref{sec:results}).

It is worth mentioning, that correlations between the various parameters in the models will artificially inflate the model uncertainties, since the multidimensional least-squares plane will have more than one unique solution. In practice, this would imply an overestimation of the confidence intervals reported here. We can, however, rule out a high degree of correlation between the various model parameters, as this would manifest itself in unstable solutions with regard to small variations in the observations. The latter is essentially probed by our various realisations of the mock input data, which show that the solutions are generally stable and recover the mock input values regardless of e.g. the noise properties.

The final modelling results are summarised in Table \ref{table:table3}, along with our set of complementary models, which we have constructed to assess uncertainties related to modelling i) a smaller FOV, ii) a miscalculation of the PSF size and iii) a single aperture measurement. Especially the latter has been carried out for direct comparison with literature measurements from \holi, which are based on a single aperture velocity dispersion.

\begin{table}
	\caption{Cosmological distance constraints from strong lensing-only and joint strong lensing and stellar dynamical models. The first column indicates the model, mock error type and S/N of the IFU stellar kinematics. Our FIDUCIAL modelling results are summarised under "Full FOV". Results for models with a smaller $2\arcsec\times2\arcsec$ FOV, an overestimated PSF and a single aperture measurement are denoted respectively. The latter three have only been modelled for a S/N of 60/40, which we deem optimal for future cosmological studies with JWST. The second column shows the constraints for the model time-delay distance \dtmod, when the combined PDF is fitted by a normal distribution with mean $\mu$ and standard deviation $\sigma$. In instances where the distribution is clearly bimodal, we quote the 50th and 16th/84th percentiles of the distribution. The third column shows the same constraints for the model lens distance \ddmod. The PDF in all cases but the gNFW is the combined and BIC weighted PDF from \texttt{COMPOSITE} and \texttt{SPEMD}s models, with 6 different source grid resolutions. In the lensing-only case, an equal weighting of all models and source resolutions is applied due to the MSD. The \textit{true} mock distances for creating the stellar kinematic maps are indicated.}
	\begin{center}
	\centerline{
	\begin{tabular}{ c  c  c }
		\hline
		\hline
		Model & \dtmod\ [Mpc] & \ddmod\ [Mpc] \\
		\hline
		\textit{True} Mock Distances & 1823  & 775 \\
		\hline
		Lensing & 1836$^{+89}_{-116}$ & \\
		\hline
		\textbf{Full FOV} & & \\
		Lensing \& Dynamics & 1762$\pm$42 & 770$\pm$18 \\
		IDEAL (30/20) & & \\
		Lensing \& Dynamics & 1767$\pm$42 & 742$\pm$17 \\
		FIDUCIAL (30/20) & & \\
		Lensing \& Dynamics & 1789$\pm$37 & 769$\pm$14 \\
		IDEAL (60/40) & & \\
		Lensing \& Dynamics & 1794$\pm$36 & 734$\pm$13 \\
		FIDUCIAL (60/40) & & \\
		Lensing \& Dynamics & 1817$\pm$30 & 776$\pm$12 \\
		IDEAL (100/60) & & \\
		Lensing \& Dynamics & 1825$\pm$30 & 748$\pm$12 \\
		FIDUCIAL (100/60) & & \\
		\hline
		\textbf{SMALL FOV} & & \\
		Lensing \& Dynamics & 1783$\pm$42 & 756$\pm$13 \\
		IDEAL (60/40) & & \\
		Lensing \& Dynamics & 1791$\pm$42 & 732$\pm$13 \\
		FIDUCIAL (60/40) & & \\
		\hline
		\textbf{PSF} & & \\
		Lensing \& Dynamics & 1790$\pm$36 & 769$\pm$14 \\
		IDEAL (60/40) & & \\
		Lensing \& Dynamics & 1794$\pm$36 & 734$\pm$13 \\
		FIDUCIAL (60/40) & & \\
		\hline
		\textbf{APERTURE} & & \\
		Lensing \& Dynamics & 1836$^{+91}_{-116}$ & 757$^{+93}_{-79}$\\
		IDEAL (60/40) & & \\
		Lensing \& Dynamics & 1836$^{+89}_{-117}$ & 726$^{+91}_{-72}$\\
		FIDUCIAL (60/40) & & \\
		\hline
		\textbf{gNFW} & & \\
		Lensing \& Dynamics & 1890$\pm$39 & 780$\pm$15 \\
		IDEAL (60/40) & & \\
		Lensing \& Dynamics & 1891$\pm$39 & 761$\pm$14 \\
		FIDUCIAL (60/40) & & \\
		\hline
	\end{tabular}
	}
%	\vspace{2ex}
	\label{table:table3}
	\end{center}
\end{table}

%---------------------------------------------------------------------
\subsection{Sources of uncertainty}
\label{sec:uncertainty}
%---------------------------------------------------------------------

To understand the impact of various sources of uncertainty, we complement our FIDUCIAL ("Full FOV") models by fitting to different data sets, by adopting different mass parameterisations and by accounting for PSF mismatches. In all cases, the data have been mocked up and modelled in the same manner as outlined in Sec. \ref{sec:stellarkinematics} and \ref{sec:setup}.

%---------------------------------------------------------------------
\subsubsection{Field Of View}
\label{sec:fieldofview}
%---------------------------------------------------------------------

In a first test, we fit to a smaller $2\arcsec\times2\arcsec$ FOV. As mentioned in Sec. \ref{sec:stellarkinematics} and \ref{sec:setup}, we have omitted the satellite and AGN images in the data construction and modelling phase. Clearly, both will result in a loss of spatial information, as the affected spaxels will have to be masked when measuring the kinematic moments. In addition to contamination from nearby sources, spatial information will also suffer due to the breakdown of a Poisson noise dominated regime in the remote regions. It is hard to quantify this loss beforehand, given that the final S/N in any given spaxel will be a complex function of the noise properties of the detector, but we try to mimic both effects by drastically reducing the FOV by almost 50\%. Keep in mind, however, that this does not translate to a loss of 50\% of spatial information. The final number of bins is still 56, compared to 78 for the nominal FOV, and a consequence of the low S/N spaxels beyond $\sim$ 1\arcsec\ being discarded, which are otherwise massively binned to reach the target S/N.

The modelling results of this run are presented in Table \ref{table:table3}, where fits to the IDEAL (i.e. with statistical noise only) and FIDUCIAL (i.e. with correlated and uncorrelated systematics) data are taken into account for a S/N configuration of 60/40. As expected, the constraints for \dtmod\ [$\mu,\sigma$]$_{\rm IDEAL}$ = [1783\,Mpc, 42\,Mpc] \& [$\mu,\sigma$]$_{\rm FIDUCIAL}$ = [1791\,Mpc, 42\,Mpc] suffer and we obtain a precision of 2.5\%, when compared to our reference "Full FOV" models with the same S/N, which has a precision of 2.1\%. While we achieve a comparable precision for \ddmod\ with this smaller FOV, it is noteworthy that the bias towards lower distances is slightly more pronounced [$\mu,\sigma$]$_{\rm FIDUCIAL}$ = [732\,Mpc, 13\,Mpc], where the \textit{true} lens distance can now only be recovered within $\sim$ 3.2$\sigma$. These findings urge us to aim for the deepest and highest S/N observations, as any loss in quality (i.e. S/N) and/or quantity (i.e. FOV) of the IFU stellar kinematics quickly diminishes any potential gain in the cosmological distance measurements. This is most evident, when we consider the extreme case of a single aperture measurement in Sec. \ref{sec:singleaperture}.

%---------------------------------------------------------------------
\subsubsection{Point Spread Function}
\label{sec:psf}
%---------------------------------------------------------------------

For the creation of the mock IFU stellar kinematics as well as during the modelling phase, we convolve the predictions of the axisymmetric Jeans equations with a PSF of 0.08\arcsec\ FWHM size. This PSF is twice as large as the diffraction limit of \jwst, but a reasonable choice considering past applications and recent simulations for the next generation of telescopes \citep{2011aoel.confE..30T}. In contrast to state-of-the-art AO assisted instruments, this remarkable angular separation is achieved by means of a significantly smaller, flux dominating PSF core. The PSF will, however, be undersampled, given NIRSpec's pixel size, and a dithering strategy is vital to achieve the nominal spatial resolution. Nonetheless, slight mismatches with the \textit{true} PSF can be expected when measuring the PSF size from real observations, and we account for this mismatch by convolving the model predictions with a PSF that is roughly 10\% larger (i.e. $\sim$ 0.09\arcsec).

Our PSF mismatch modelling results are again summarised in Table \ref{table:table3}. In light of the sub-pixel PSF size, the cosmological distance constraints are stable across both error assumptions (i.e. for statistical noise only and with correlated and uncorrelated errors on top), yielding almost identical precision on both \dtmod\ and \ddmod. We therefore omit probing models of different S/N and note that our constraints are insensitive to minor deviations from the \textit{true} PSF size. Keep also in mind, that the correlation between individual bins is minimal. Our PSF is of the order of \jwst\ NIRSpec's nominal pixel size of 0.1\arcsec. Besides the most central bins, which consist of individual spaxels, most kinematic measurements are essentially independent, given that they are also considerably larger, even if we assume a PSF which is 3$\times$ as large as the diffraction limit.

%---------------------------------------------------------------------
\subsubsection{Single aperture}
\label{sec:singleaperture}
%---------------------------------------------------------------------

Stellar kinematics are now commonly employed to break the inherent modelling degeneracies in (time-delay) strong lensing studies \citep[e.g.][]{2002ApJ...575...87T,2003ApJ...599...70K,2004ApJ...611..739T,2006ApJ...649..599K}. Yet, due to the faintness of the lens and difficulties in separating the bright quasar light from the galaxy, even with state-of-the-art facilities, the data is confined to a single aperture measurement of the stellar velocity dispersion. Moreover, currently employed techniques utilising this kinematic information in strong lensing studies are usually not self-consistent or physically too simple to capture the true complexity of realistic lens galaxies. For instance, most implementations in strong lensing studies assume elliptical lens mass models, but model predictions for the stellar kinematics are based upon a spherically symmetric mass distribution \citep[e.g.][]{2019MNRAS.484.4726B,2015JCAP...11..033J,2013ApJ...766...70S}. Similarly, simple assumptions for the velocity anisotropy profile are made \citep{1979PAZh....5...77O,1985MNRAS.214P..25M,1985AJ.....90.1027M}.

In order to assess the impact of using a single aperture measurement on \dtmod and \ddmod, under the aforementioned modelling limitations, we mock up stellar kinematics of \rxj\ within a 0.8\arcsec$ \times$ 0.8\arcsec\ FOV at \jwst\ resolution, according to the procedure outlined in Sec. \ref{sec:stellarkinematics}. The kinematics are then luminosity weighted to simulate a single aperture measurement of $\overline{v_{\mathrm{LOS}}^{2}}$. In principle, we would have to split $\overline{v_{\mathrm{LOS}}^{2}}$ into two first order moments \citep{1980PASJ...32...41S}, defining the contribution of its ordered vs. random motions, for a straightforward comparison with a stellar velocity dispersion measurement. For simplicity, however, we assume that this massive elliptical lens is dispersion dominated such that $\overline{v_{\mathrm{LOS}}^{2}} \approx \sigma^{2}$ within our FOV.

Contrary to our previous models in this section, we now break the self-consistency of our joint lensing and stellar dynamical models. We run the axisymmetric Jeans models in the spherical limit, by fixing the projected short- vs. long-axis ratio of the luminous and dark matter distribution to $q' = 0.99$. This is in line with literature studies, where spherical Jeans models have been employed for fitting the stellar velocity dispersion. For both the \texttt{COMPOSITE} and \texttt{SPEMD}, the models can easily recover the single aperture \vlos\ measurement of 325$\pm$12 \kms, yielding very similar goodness of fit values across all source resolutions. Ergo, the BIC is not capable of discerning between the two different mass parameterisations, leaving the final distribution still double peaked with \dtmod$ = 1836^{+89}_{-117}$\,Mpc for our FIDUCIAL models (median and 16th \& 84th percentiles). The consequences of anchoring the cosmological distance measurements in \rxj\ on a single aperture velocity dispersion are most noticeable for \ddmod, where the 1D PDF is only loosely constrained, implying a precision $\ge 10\%$. This is smaller than the 18\% found in \cite{2019Sci...365.1134J}, based on spherical Jeans models that fit the literature stellar velocity dispersion of 323$\pm$20 \kms, but the difference is likely to be attributed to the smaller errors in our mock data. Even though the precision on the distance measurements degrades substantially with only a single aperture averaged second order velocity moment instead of a 2D kinematic map, the input distances are recovered well within 1$\sigma$, without obvious signs of bias despite the spherical symmetry assumption employed here. This seems to be counterintuitive when compared to our elliptical mass models (see Table \ref{table:table3}), which fit the 2D kinematics and can exhibit strong biases (especially in \dd), but can be explained by two effects. First, the single aperture measurement is simply a tracer of the enclosed mass, in contrast to the 2D kinematics, which traces the spatially resolved mass distribution in detail. Both the COMPOSITE and SPEMD are capable of recovering the enclosed mass within the errors, with no preference for either model. Given the relatively loose constraints from the single anchor measurement in the single aperture case, the corresponding errors in the cosmological distance inferences from a single lens are sufficiently large to cover any potential bias. With more lenses and thus a reduction in the uncertainty of global parameters (e.g. $H_0$), such a bias could then become significant relative to the uncertainty and lifting the spherical assumption for the kinematics (as we have done previously) becomes important.  Secondly, \rxj\ is quite roundish with a mean flattening for the light profile and NFW halo of $\sim$0.85 and $\sim$0.75 respectively. Recovering the input kinematics with axisymmetric Jeans models in the spherically symmetric case is therefore feasible without introducing a significant bias. This might change for highly flattened gravitational lenses, which (however) are less likely to be found, given the fact that the lensing cross section increases with galaxy mass and the most massive galaxies being quite roundish in projection.

We also emphasise that our assumption of a dispersion dominated lens within the single aperture FOV is unlikely to introduce a significant bias in our inference. With any net streaming motions present, our mock $\vrms$ data cannot be attributed to $\sigma$ only. As a consequence, the true velocity dispersion must be lower. Assuming, however, net streaming motions of e.g. 50 \kms\ within the 0.8\arcsec $\times$ 0.8\arcsec\ FOV\footnote{We assume low net streaming motions, since i) the aperture FOV only covers half the half-light radius of \rxj, ii) the contributions of $v$ are luminosity weighted, such that the contribution of the higher velocity wings will be downweighted, and iii) the increase of the lensing cross section increases with galaxy mass and the fact that the most massive galaxies are found to be slow-rotating objects, with low $v/\sigma$ profiles \citep[e.g.][]{2007MNRAS.379..418C}.}, results in a change of only 4 \kms\ for $\sigma$. Considering our mock $\vrms$ observations of 325$\pm$12 \kms, the change in $\sigma$ is thus well within the error bars. We therefore conclude that our findings for models which employ single aperture kinematics are likely to hold,  unless the lens system exhibits significant rotation or is highly flattened. It is worth noting, however, that a single aperture measurement is less powerful in lifting the MSD as a result, due to the minuscule differences also in the kinematic $\chi^2$.

%---------------------------------------------------------------------
\subsubsection{Generalised NFW}
\label{sec:gnfw}
%---------------------------------------------------------------------

In a last effort to quantify the systematic uncertainties in our models, we adopt a generalised NFW profile \citep[\texttt{gNFW,}][]{1996MNRAS.278..488Z}, where the halo follows a density distribution according to
\begin{equation}
\label{eqn:eqn42}
\rho(r) = \frac{\rho_{0}}{(\frac{r}{r_{s}})^\gamma\;(\frac{1+r}{r_{s}})^{3-\gamma}},
\end{equation}
with $\rho_{0} = \delta_{c}\;\rho_{c}$ being a product of the characteristic density $\delta_{c}$ and the critical density $\rho_{c} = 3 H^2/8\pi G$ at the time of halo formation, halo scale radius $r_{s}$ and density slope $\gamma$. The use of a \texttt{gNFW} halo is physically motivated by dissipational cosmological simulations, where the dark halo reacts to an accumulation of the central baryonic component via contraction \citep{1986ApJ...301...27B}. More importantly, though, a mass model with a halo of \texttt{gNFW} form will allow us to better understand the systematics associated with a mass model which is comparably close to the \textit{true} lens mass distribution. This is particularly interesting, given our general ignorance of the \textit{true} underlying mass distribution. To this end, we make use of a \texttt{gNFW} halo with a Gaussian prior on the halo slope $\gamma$ with mean and standard deviation $1.0\pm0.1$ \citep{2016MNRAS.462..681M}, while adopting the same priors for the remaining variables.

When employing the above density parameterisation for the dark halo, we obtain strong biases for the time-delay distance, whereas the lens distance again is susceptible to any systematics in the kinematic measurements. With \dtmod\ [$\mu,\sigma$]$_{\rm IDEAL}$ = [1890\,Mpc, 39\,Mpc] \& [$\mu,\sigma$]$_{\rm FIDUCIAL}$ = [1891\,Mpc, 39\,Mpc]  and \ddmod\ [$\mu,\sigma$]$_{\rm IDEAL}$ = [780\,Mpc, 15\,Mpc] \& [$\mu,\sigma$]$_{\rm FIDUCIAL}$ = [761\,Mpc, 14\,Mpc], we achieve a 2.1\% and 1.9\% precision measurement for \dtmod\ and \ddmod\ respectively. However, the accuracy for the time-delay distance measurement suffers significantly, with \dtmod\ only being recovered within $2\sigma$. This simple toy model is another display of the MSD, where comparably good fits to the lensing data are obtained while yielding significantly different distance constraints. Only by means of the BIC due to i) slightly worse kinematic likelihoods of these models and ii) an increase in the model degrees of freedom\footnote{Despite our Gaussian prior with mean $\gamma=1.0$, the models strongly converge towards a much shallower density slope of $\gamma \sim 0.6$. Models with a flat prior of $0.5 \le \gamma \le 1.5$ yielded identical constraints. In contrast to the scale radius $r_s$, where we used a Gaussian prior as it is not constrained by data, the halo slope is constrained and therefore constitutes an additional degree of freedom in the \texttt{gNFW} models.}, can we break this degeneracy. Even if the contribution of the \texttt{gNFW} models to the final distance constraints is non-negligible, in contrast to the \texttt{SPEMD} mass models which have vanishing posterior weights (see Appendix \ref{sec:appendix}), the final cosmological inference (see Sec. \ref{sec:forecast}) is still vastly improved when compared to the literature single aperture findings.\\

%---------------------------------------------------------------------
\subsubsection{Other sources of uncertainty}
\label{sec:other_uncertainties}
%---------------------------------------------------------------------

Optimising the likelihood function in Eq. \ref{eqn:eqn28}, implicitly assumes uncorrelated measurements in the lensing and kinematic data. While this is a sensible assumption for the surface brightness and line-of-sight velocity moments, the time delays between individual pairs of images are correlated to some degree. To test for the impact of covariances between the image pairs, we adopt and include the covariance matrix of the time-delay errors into our models, as measured and reported in \citep{2013A&A...556A..22T}. From a single test, based on our \texttt{COMPOSITE} mass model with a source resolution of $64\times64$ pixels and fitting to our FIDUCIAL data, we find no differences ($< 1\%$) for the inferred cosmological distances \dt\ and \dd, which is not surprising given the weak correlations in the measurement errors between individual image pairs \citep[][Fig. 8]{2013A&A...556A..22T}.\\

When constructing mass models of increased complexity, we have knowingly omitted to probe models with e.g. a radially varying stellar $M/L$. Despite mounting evidence for such an IMF induced change \citep[see e.g.][]{2010Natur.468..940V,2015MNRAS.447.1033M,2017ApJ...837..166C}, our dark halo already already incorporates radial variations in the total $M/L$, albeit with a different slope than the baryonic component. That is, the increased model degrees of freedom in the \texttt{gNFW} profile allow us to assess the implications for the cosmological distances, if the $M/L$ profile of the model differs from the data and vice versa. Considering the findings in Sec. \ref{sec:gnfw}, even the slightest mismatches in the $M/L$ could have adverse consequences for the cosmological distance measurements, while providing acceptable fits to the lensing-only data. However, assuming that the family of mass models include the \textit{true} lens model (as is the case in this study), these will be properly downweighted by the BIC. In any case, this convincingly demonstrates the paramount importance of adopting a wide range of plausible mass parameterisations for TDC purposes, 
%SHS19-1102% (including radial variations in the stellar M/L), 
as strong offsets from the \textit{true} time-delay distance will be measured otherwise, even if high-quality IFU kinematics are included in the fit.

%============================= Section 5 =============================
\section{Cosmological Forecast \& Discussion}
\label{sec:results}
%=====================================================================

We describe the cosmological constraints in flat $\Lambda$CDM using the forecasted distance measurements from the last section for both the IDEAL and FIDUCIAL lensing and dynamics models with a S/N of 60/40. Our final inference is based on the "Full FOV", "PSF" and "gNFW" models, given that these make use of the same data sets and allow for a proper evaluation and comparison within the BIC. While the former two models have been probed by both the \texttt{COMPOSITE} and \texttt{SPEMD} mass models, the latter employs a generalised NFW model, which allows us to include a wide range of systematic effects related also to the lens mass parameterisation. We also compare the constraints from using the joint $\dt$-$\dd$ measurement, with that from using only the marginalised $\dt$ measurement.

%---------------------------------------------------------------------
\subsection{Importance sampling with the forecasted distances}
\label{sec:sampling}
%---------------------------------------------------------------------

In order to obtain a cosmographic forecast, the first step is to get the posterior probability distribution of the cosmological distance measurements, accounting for systematic uncertainties.  From the lensing and dynamical modelling detailed in the previous section, we have Markov chains containing the sampled \dtmod\ and \ddmod\ parameters, for various models and set ups.  For each data set, we weight the various models (with different lensing mass parametrisation and lensing source grid resolutions) using their BIC values following Eq. \ref{eqn:eqn40}, where we have estimated $\sigma_{\rm BIC}$ through the scatter in BIC values from models that differ only in lensing source resolutions (given that the source resolutions have a dominant effect on the scatter).  We then combine the weighted chains/models, and fit the marginalised \dtmod-\ddmod\ distribution with a multivariate Gaussian to obtain $P(\dtmod, \ddmod | d_{\mathrm{L}}, d_{\mathrm{D}})$.

To further account for the uncertainty due to the external convergence from mass structures along the LOS, we use the $\kext$ distribution from \citet{2017ApJ...836..141M}, which is obtained through a 3-dimensional reconstruction of the mass structures in the field of \rxj. By reconstructing the mass distribution specific to the \rxj\ sightline, the resulting $\kext$ distribution is substantially more precise compared to that obtained statistically through galaxy number counts and numerical simulations in \cite{2014ApJ...788L..35S}.  While the galaxy-number counts approach is thoroughly tested and robust \citep[e.g.,][]{2013ApJ...768...39G, 2017MNRAS.467.4220R}, there is often a multitudes of sightlines with similar galaxy-number counts with different $\kext$, giving rise to the scatter/uncertainty in $\kext$ from this approach.  By focusing on the environment of specific lens, the new method developed by \citet{2017ApJ...836..141M} \citep[see also ][]{2013MNRAS.432..679C} help reduce the scatter and uncertainty in $\kext$, and we expect further developments/applications of this new method to yield better constrained $\kext$ for lens systems in the future.  In this paper, we adopt and approximate the $\kext$ from \cite{2017ApJ...836..141M} as a Gaussian distribution with mean 0.05 and standard deviation of 0.01 (McCully, private comm.), and we also consider the $\kext$ from the number counts \citep{2014ApJ...788L..35S} for comparison.
%We draw a value of $\kext$ from the distribution for each sample in the Markov chains of our lensing and dynamical modelling, and scale the \dtmod\ and \ddmod\ parameters in the chain according to Eqs. (\ref{eqn:eqn13}) and (\ref{eqn:eqn35}) to obtain $\dt$ and $\dd$.  We then fit a multi-variate Gaussian to the marginalised \dt-\dd\ distribution, to obtain $P(\dt, \dd | d_{\mathrm{L}}, d_{\mathrm{D}}, d_{\mathrm{env}})$, where $d_{\rm env}$ denotes the environment data used to obtain $\kext$.

%$P_i(\dt, \dd | d_{\mathrm{L}}, d_{\mathrm{D}}, d_{\mathrm{env}})$, where $i$ is \texttt{SPEMD} or \texttt{COMPOSITE}, and $d_{\rm env}$ denotes the environment data used to obtain $\kext$.  Since these two models yield comparable goodness of fit (a manifestation of the mass-sheet degeneracy) and are both equally plausible models from previous studies, we weight the two models equally \todo{Sherry}{is this still true, since we're using the BIC to weight properly now?!} to obtain a conservative estimate of the distance constraints:
%\begin{equation}
%\label{eqn:P_DtDd}
%\begin{split}
%P(\dt, \dd | d_{\mathrm{L}}, d_{\mathrm{D}}, d_{\mathrm{env}}) = P_{\mathrm{\texttt{SPEMD}}}(\dt, \dd | d_{\mathrm{L}}, d_{\mathrm{D}}, d_{\mathrm{env}}) \\ + P_{\mathrm{\texttt{COMPOSITE}}}(\dt, \dd | d_{\mathrm{L}}, d_{\mathrm{D}}, d_{\mathrm{env}}).
%\end{split}
%\end{equation}

With the posterior probability distribution $P(\dtmod, \ddmod | d_{\mathrm{L}}, d_{\mathrm{D}})$, we can relate this to constraints on the cosmological parameters in any background cosmology through importance sampling \citep[e.g.,][]{2002PhRvD..66j3511L, 2010ApJ...711..201S}.  As an illustration, we consider the constraints on $H_0$ specifically for the flat \lcdm\ cosmology, where we adopt uniform priors on \h\ between [50,120]\,\kmsM and on the matter density parameter
$\Omega_{\mathrm{m}}$ between $[0.05,0.5]$.  We draw $10^7$ samples in $\{H_0, \Omega_{\mathrm{m}}\}$, and compute the corresponding $\dt$ and $\dd$ values given the lens and source redshifts in flat \lcdm.  For each of these samples, we also draw a value of $\kext$ from the $\kext$ distribution, and scale the distances according to Eqs. \ref{eqn:eqn13} and \ref{eqn:eqn33} to obtain $\dtmod$ and $\ddmod$.  We finally weight the sample by $P(\dtmod, \ddmod | d_{\mathrm{L}}, d_{\mathrm{D}}$). From the distribution of the weighted samples, we obtain constraints on $H_0$ and $\Omega_{\mathrm{m}}$.

%---------------------------------------------------------------------
\subsection{Forecasted $H_0$ constraint in flat $\Lambda$CDM}
\label{sec:forecast}
%---------------------------------------------------------------------

\begin{figure*}
\begin{center}
\includegraphics[width=0.9\linewidth]{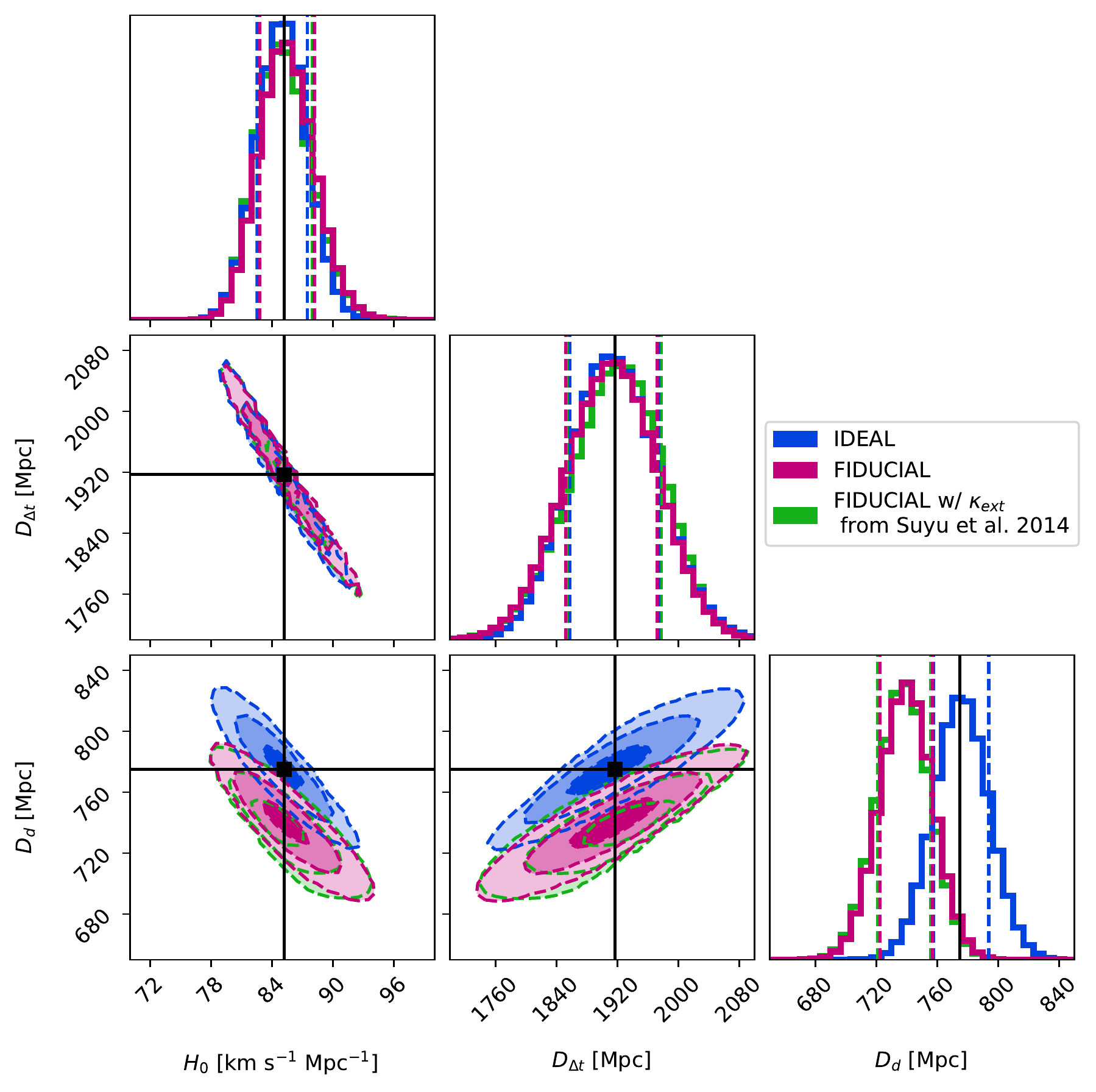}
\end{center}
\caption{Cosmological constraints from our joint strong lensing and stellar dynamical models of \rxj, for flat \lcdm\ with uniform priors on \h\ of [50,120]\,\kmsM and $\Omega_{\mathrm{m}}$ of [0.05,0.5]. The magenta shaded contours show the 1, 2 and 3$\sigma$ confidence intervals for our FIDUCIAL models with S/N of 60/40 (i.e. including correlated and uncorrelated systematic errors in the IFU stellar kinematics). The black points (lines) depict the mock input values after accounting for a \kext\ distribution with mean 0.05 and standard deviation of 0.01 \citep{2017ApJ...836..141M}. The blue shaded contours show the corresponding constraints for our models with statistical noise only (i.e. without correlated and uncorrelated systematic errors). The green shaded contours are obtained from our FIDUCIAL models with a S/N of 60/40 when $\kext$ is estimated from number counts along overdense lens LOSs \citep{2014ApJ...788L..35S}. Both \h\ and \dt\ are recovered incredibly well in our FIDUCIAL models, with a precision of 3.2\% and 3.1\% respectively. While we can quote a precision of 2.4\% on \dd, the recovered value is highly biased towards lower distances due to the systematic floor we have added to \vlos, in order to mock real observational errors.}
\label{fig:fig8}
\end{figure*}

We show in Fig. \ref{fig:fig8} the cosmographic constraints for the FIDUCIAL lensing and dynamical models with S/N of 60/40 (magenta). The constraints with statistical errors only (blue) are comparable, though slightly shifted in \dd\ towards the mock input value of 775\,Mpc. After including all sources of uncertainty, we expect to achieve a measurement of \h\ $= 85.4^{+2.8}_{-2.6}$ \kmsM, with 3.2\% precision (defined by the 50th, 16th and 84th percentile), by having high-quality spatially-resolved kinematic data from JWST. The marginalised $\dt$ constraint is $1913^{+59}_{-61}{\rm Mpc}$, which is of similar precision as $H_0$.  Compared to the 6.6\% uncertainty in $\dt$ without spatially resolved kinematic data \citep{2014ApJ...788L..35S}, we are reducing the systematic uncertainty by a factor of $\sim$ 2. Even when accounting for a more conservative \kext\ distribution \citep{2014ApJ...788L..35S}, by ray tracing through overdense LOSs in the Millennium Simulations \citep{2005Natur.435..629S}, a measurement of \h\ $= 85.1^{+2.9}_{-2.6}$ \kmsM\ (i.e. with 3.3\% precision) is within reach, partly constrained by the assumption of flat $\Lambda$CDM, which restricts the range of plausible $\dt$ and $\dd$ values. That is, most of our improved constraints (a factor of $\sim$2 in precision or 3.3\% respectively) stem from our mock 2D \jwst\ kinematics, when compared to the lensing-only case, with an additional improvement of 0.1\%, if the smaller scatter in the LOS convergence \citep{2017ApJ...836..141M} is taken into account.

\begin{figure}
\begin{center}
\includegraphics[width=0.99\linewidth]{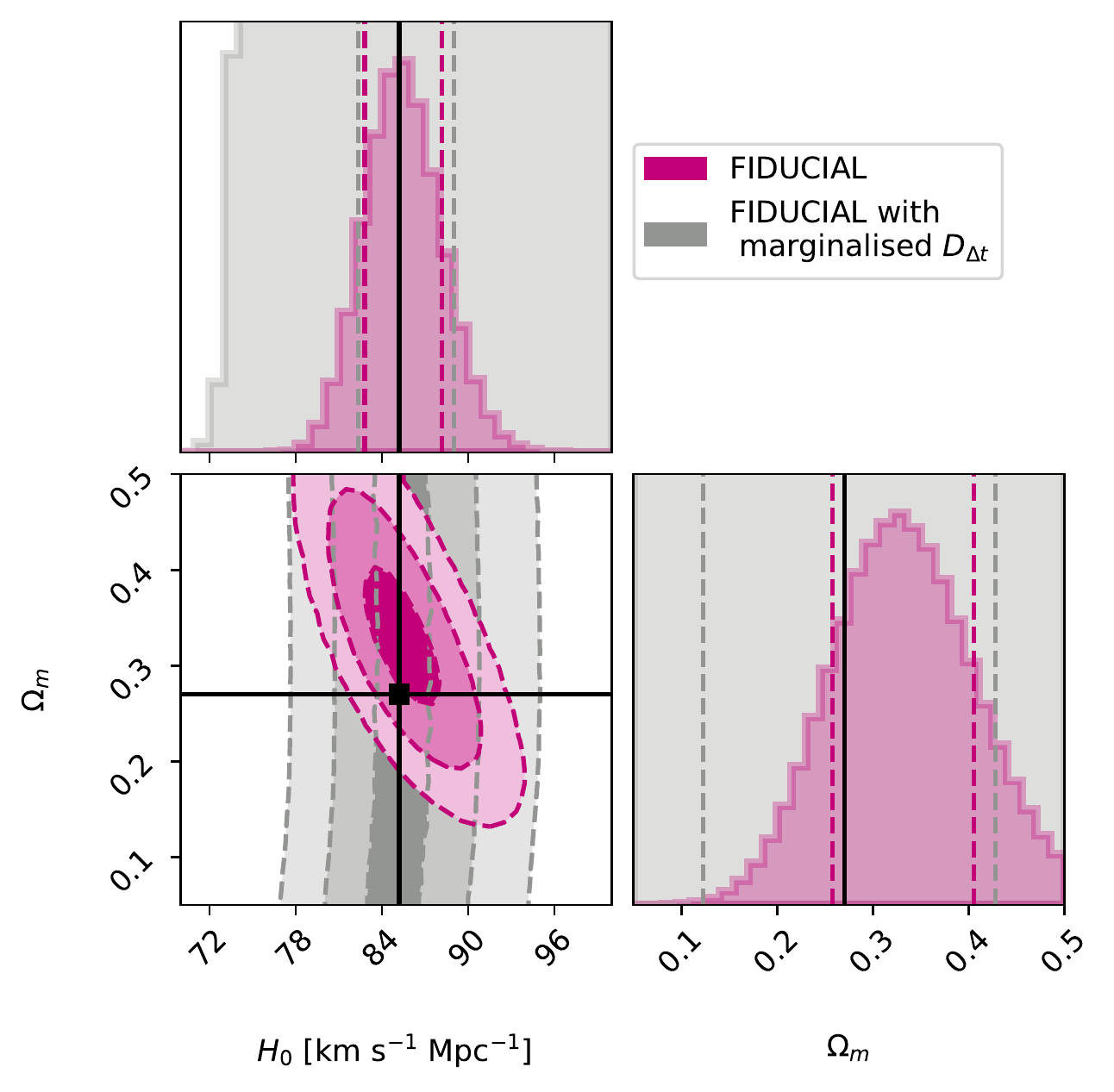}
\end{center}
\caption{Cosmological constraints from our joint strong lensing and stellar dynamical models of \rxj, for flat \lcdm\ with uniform priors on \h\ of [50,120]\,\kmsM\ and $\Omega_{\mathrm{m}}$ of [0.05,0.5]. The magenta shaded contours show the 1, 2 and 3$\sigma$ confidence intervals for our FIDUCIAL models with S/N of 60/40 (i.e. including correlated and uncorrelated systematic errors in the IFU stellar kinematics). The grey shaded contours show the corresponding constraints for the marginalised \dt\ models, i.e. effectively for lensing-only models. The different tilts in \h$-\Omega_{\mathrm{m}}$, when marginalised over \dt\ and \dd\ respectively, break some of the degeneracies.  The tight measurement of \dd\ from 2D kinematics complements the \dt\ measurement and leads to much improved constraints for the matter density.}
\label{fig:fig9}
\end{figure}

\begin{figure*}
\begin{center}
\includegraphics[width=0.9\linewidth]{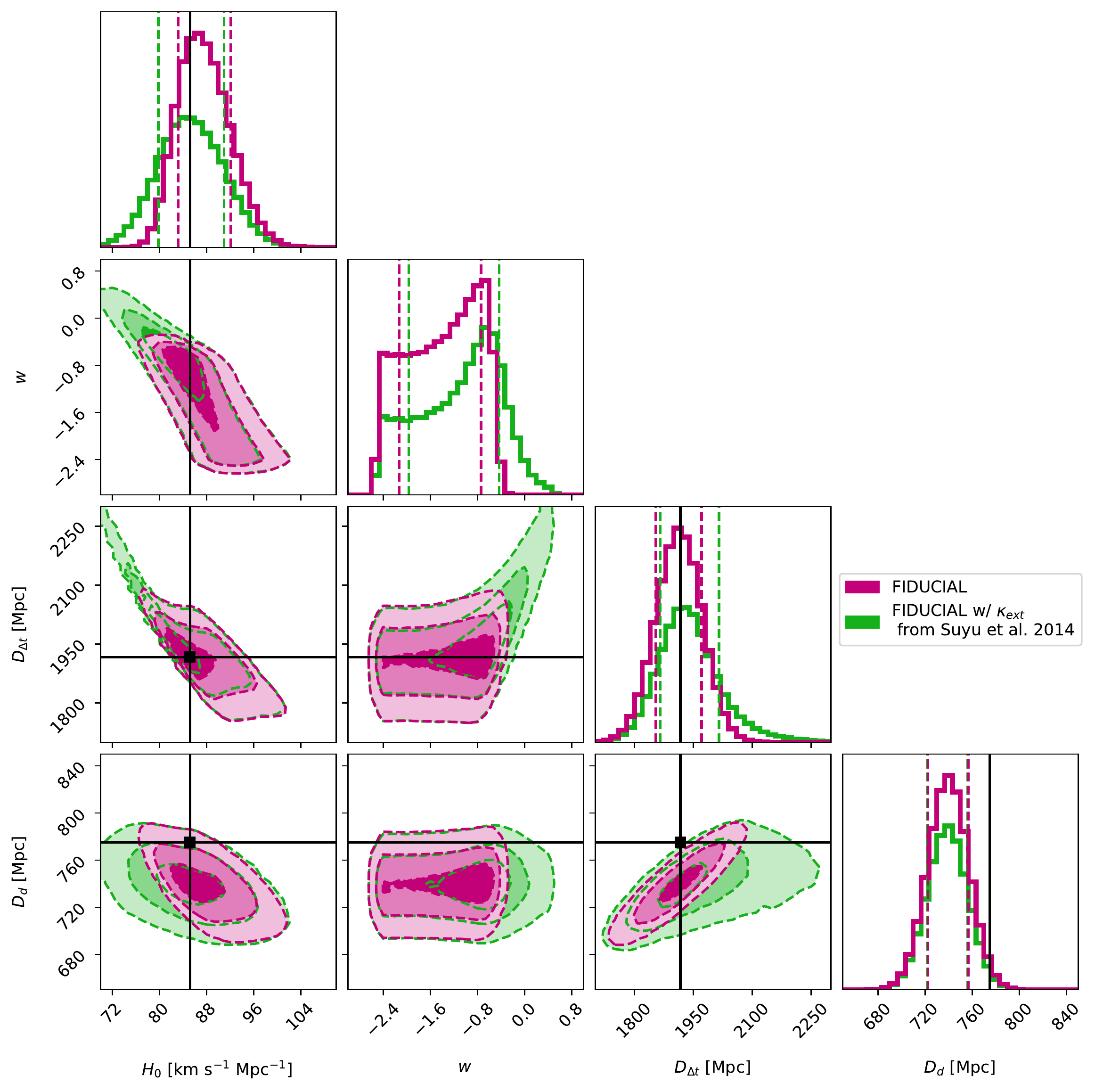}
\end{center}
\caption{Cosmological constraints from our joint strong lensing and stellar dynamical models of \rxj, for flat $w$CDM with uniform priors on \h\ of [50,120]\,\kmsM, $\Omega_{\mathrm{m}}$ of [0.05,0.5] and $w$ of [$-2.5$,0.5]. The magenta shaded contours show the 1, 2 and 3$\sigma$ confidence intervals for our FIDUCIAL models with S/N of 60/40 (i.e. including correlated and uncorrelated systematic errors in the IFU stellar kinematics). The black points (lines) depict the mock input values after accounting for a \kext\ distribution with mean 0.05 and standard deviation of 0.01 \citep{2017ApJ...836..141M}. The green shaded contours are obtained from our FIDUCIAL models with a S/N of 60/40 when $\kext$ is estimated from number counts along overdense lens LOSs \citep{2014ApJ...788L..35S}. Even if $w$ is only loosely constrained, given that \dt\ is mainly sensitive to \h, the constraints from this single lens system yield similar precision as literature studies of 3 lenses without IFU kinematics and provides a promising avenue for the exploration of models beyond \lcdm.}
\label{fig:fig10}
\end{figure*}

With spatially resolved kinematics, we would also constrain $\dd$ to $740^{+17}_{-18}$, with 2.4\% uncertainty. This measurement is substantially better than the $\sim 18\%$ from the single-aperture average velocity dispersion measurement \citep{2015JCAP...11..033J,2019Sci...365.1134J}, yet biased towards lower values due to the systematic floor we have added to the FIDUCIAL mock kinematics. Nonetheless, we recover the mock input lens distance within $\sim 2\sigma$. To see whether $\dd$ helps to further constrain $H_0$, we repeat the cosmographic forecast above using the marginalised probability distribution of $\dt$, i.e., $P(\dt | d_{\mathrm{L}}, d_{\mathrm{D}}, d_{\mathrm{env}}) = \int {\rm d}{\dd} P(\dt, \dd | d_{\mathrm{L}}, d_{\mathrm{D}}, d_{\mathrm{env}})$.  With only the marginalised $\dt$, the constraint on $H_0$ degrades slightly to $85.5^{+3.5}_{-3.3}$ (i.e. 4.0\% uncertainty) for the FIDUCIAL model. Note, however, that the marginalised \dt\ measurement is not totally ignoring the distance information in \dd, as it has been used to break the MSD in the first place. Moreover, the combination of two distance measurements from a single lens provides not only tight constraints on \h, but adds significant constraining power also for the matter density $\Omega_\mathrm{m}$ (Fig. \ref{fig:fig10}).

Given the low redshift of the lens galaxy in \rxj, $H_0$ is primarily constrained by $\dt$ in this case, but the tight constraint on $\dd$ would provide substantial constraints on cosmological models beyond flat $\Lambda$CDM. For illustration, we show the cosmographic constraints for flat $w$CDM cosmology in Fig. \ref{fig:fig10}, where  we focus on our FIDUCIAL models with improved \citep{2017ApJ...836..141M} and conservative \citep{2014ApJ...788L..35S} \kext\ distributions. We note that the prior range on $\dt$ and $\dd$ in the more general $w$CDM model is substantially broader compared to flat $\Lambda$CDM; the conservative $\kext$ distribution thus leads to a wider $\dt$ distribution, in comparison to the case of $\Lambda$CDM in Fig.~\ref{fig:fig8}.  With $w = -1.35^{+0.61}_{-0.77}$, the time-independent dark energy is only loosely constrained but, nonetheless, comparable to the combined constraints from 3 single lenses without 2D kinematic data. \citep{2017MNRAS.465.4914B}. Also, such a $\dd$ measurement would serve as a stringent anchor for the inverse distance ladder approach for inferring $H_0$ \citep{2019Sci...365.1134J}.

Our forecasted constraints are essentially limited by the S/N of the spatially resolved kinematic maps for breaking lens model degeneracies. As shown in the previous section, higher S/N helps to discriminate between the different mass parameterisations better and hence provide tighter constraints on \dt, as the differences in the goodness of fit (and thus in the BIC weighting) become more prominent. For the case of S/N of 60/40 of the FIDUCIAL model, the resulting $H_0$ would have an uncertainty of 2.4\%. While this S/N can be achieved with reasonable observation times ($\sim$\,6h), higher quality data to constrain \h\ further would be difficult to obtain with JWST given the long integration time needed. Future giant segmented mirror telescopes
like the \elt\ and \tmt, however, could achieve a S/N $\ge$ 100/60 within the same time, owing to their $\sim 5-6\times$ larger aperture.

%============================= Section 6 =============================
\section{Summary \& Outlook}
\label{sec:summary}
%=====================================================================

In this paper, we presented a self-consistent joint strong lensing \& stellar dynamical modelling machinery for TDC purposes, which employs a pixelated source reconstruction model and the solutions of the Jeans
equations in axial symmetry. Our analysis is carried out within the framework of Bayesian statistics and suited, especially, for the study of strong lens configurations for which IFU stellar kinematic data
will become available in the near future, by means of the next generation of ground- and space-based telescopes. To assess the performance of the machinery and the expected gain in the inference of
cosmological distances and parameters, we mocked up IFU observations of the prominent lens system \rxj, at \jwst\ NIRSpec resolution. \rxj\ was a particularly natural choice for this study as it is the
brightest known lens galaxy for which precise time-delay measurements are already available. The mock stellar kinematic map was based on the best-fitting lensing-only mass model, with a dark and luminous matter
contribution, while making random assumptions about the orbital anisotropy and viewing orientation. The mock lens distance has been obtained by assuming a standard cosmological model with \h\ $ = 82.5$
\kmsM, $\Omega_{\mathrm{m}} = 0.27$, $\Omega_{\mathrm{\Lambda}} = 0.73$ and a lens and source redshift of $z_{\mathrm{d}} = 0.295$ and $z_{\mathrm{s}} = 0.654$, respectively.

With this suite of data, consisting of deep \hst\ imaging, precise time-delays and mock IFU stellar kinematics of various levels of quality and including various sources of uncertainty, we constructed joint strong lensing \& stellar dynamical models. Our models relied on two different mass parameterisations (\texttt{COMPOSITE} and \texttt{SPEMD}), which have been shown to yield significantly different time-delay distances when lensing-only fits are carried out \citep{2014ApJ...788L..35S}. Given the vast amount of information from the spatially resolved kinematics,
we utilised the systematic differences in the predicted second-order LOS velocities between the different models, to apply a model selection according to the BIC. The main results of our study can be
summarised as follows:
\begin{itemize}
\item The models recover remarkably well our input time-delay distance \dt\ ($\le 1\sigma$), when high-quality IFU stellar kinematics (S/N $\ge$ 60/40) are available. This result is irrespective of the IFU stellar kinematic errors (i.e. assuming purely statistical errors or with correlated and uncorrelated systematics of 2\% each included).
\item The time-delay distance can only be recovered within $1.2\sigma$ or worse, when the S/N of the IFU kinematics degrades below that of our reference 60/40 model.
\item The lens distance \dd\ is recovered in all cases within $1\sigma$, when purely statistical errors for the stellar kinematics are assumed. But, a strong offset from the mock lens distance is observed, when the stellar kinematics are systematically biased towards higher or lower values. In these instances, the \textit{true} lens distance can only be recovered within $3\sigma$ or worse, depending on the systematic offset of the data. Controlling the systematics in the measurement of the stellar kinematics is therefore key for a reliable inference of \dd.
\item The aforementioned results are valid for a 2D map, that covers the LOS velocity distribution within a 3\arcsec$\times$3\arcsec\ FOV (e.g. \jwst\ NIRSpec nominal FOV). Modelling a smaller 2\arcsec$\times$2\arcsec\ FOV, to account for loss of spatial information due to contamination from nearby objects, yields similar accuracy but is less precise ($\Delta\dtmod/\dtmod = 2.5\%$). This highlights the importance of deep and high-quality IFU data, as the gain in the cosmological inference is easily diminished when less spatial information is available.
\item Small mismatches with the \textit{true} kinematic PSF size have a negligible impact on the final modelling constraints.
\item A single aperture stellar velocity dispersion is not very effective in breaking the MSD in \rxj, yielding a marginal improvement in precision for \dtmod\ over lensing-only models. The constraints for \ddmod\ suffer the most, with a precision $\ge 11\%$.
\item We achieve a 2.0\% precision measurement on \dtmod, for our FIDUCIAL models with a S/N of 60/40 and including various sources of uncertainty while mocking up the IFU stellar kinematics. Accounting for a wide range of additional sources of systematic uncertainties, as sampled by our test models in Sec. \ref{sec:modelling}, this translates to a 3.1\% and 3.2\% precision measurement on \dt\ and \h, respectively, in flat \lcdm.
\item A 2.4\% precision measurement can be achieved for \dd. Yet, this measurement is sensitive to the aforementioned systematics in the stellar kinematics, since the constraints are mainly anchored by the IFU data.
\item The constraints for \dt, \dd\ and hence \h\ improve by a factor of $\sim$ 2, when high-quality IFU stellar kinematics are incorporated in the fit. The improvement can be traced back to three effects in particular, i) a smaller width of the PDF for individual mass models with different source resolutions, ii) a shift of the mean of the distribution towards the \textit{true} time-delay distance and iii) a drastic downweighting of models with a significantly worse goodness of fit, which is otherwise not feasible due to the MSD.
\end{itemize}

The increased flexibility of our models allows for a more realistic modelling approach, while circumventing many of the assumptions and limitations of literature time-delay studies. Yet, as the lensing cross section increases with mass, gravitational lenses are likely to be massive elliptical galaxies, which have grown through numerous violent minor and major merger encounters \citep{2016MNRAS.456.1030W}. As a consequence, lens galaxies are neither spherical nor elliptical. In fact, recent studies strongly indicate that the most massive galaxies are triaxial \citep{2018ApJ...863L..19L}, and modelling within an axisymmetric framework might be equally inadequate. It is beyond the scope of this paper to quantify the systematic uncertainties that can be traced back to the violation of axial symmetry, but literature studies show that the reconstructed dynamical masses can be underestimated by as much as 50\%, depending on the viewing orientation \citep{2007MNRAS.381.1672T}. Taking into account the link between the gravitational potential and the excess time delays, we advise against a modelling within this framework if strong signatures of triaxiality, such as isophotal twists or kinematically decoupled components, are present.

The modelling machinery presented in this paper, along with high-quality IFU data from future space- and ground-based telescopes, provides a promising outlook for constraining cosmological parameters to the few percent level from axisymmetric lenses. Given our forecast for the single lens system \rxj, an \h\ measurement of $\le$ 2.0\% precision could be within reach, if similar gains in precision can be obtained for a total of three lens systems. This would be an important boost in precision when compared to the combination of three such lens systems without 2D kinematic data \citep{2017MNRAS.465.4914B} and comparable to the current best cosmological probes. It is worth noting, however, that the stellar kinematics have been mocked up by means of a \texttt{COMPOSITE} mass model, which in turn was used to model the suite of data. In reality, though, the set of candidate models is unlikely to contain the true form of the lens potential. As the BIC only applies a relative weighting scheme between all available models, the final accuracy and precision of the cosmological inference heavily relies on an adequate description of the true lens potential. In fact, even the slightest deviations from the true lens potential, as demonstrated by our \texttt{gNFW} models, can result in a biased inference of \h, which is in agreement with \cite{2018MNRAS.474.4648S}, where a simple power law model was found to be insufficient to provide an unbiased measurement of \h\ in most cases. As a consequence, the study presented here can only be regarded as a best-case scenario, and we strongly encourage to probe a large set of flexible and physically motivated lens mass parameterisations with sufficient degrees of freedom in the radial density profile, to minimise the systematic errors associated with it. 

%============================= Acknowledgements =============================
\section*{Acknowledgements}

We thank G.~Chiriv{\`i}, T.~Treu, E.~Komatsu and S.~Birrer for helpful discussions and comments on the manuscript, and C.~McCully for sharing with us the $\kappa_{\rm ext}$ distribution from \cite{2017ApJ...836..141M}.  AY and SHS thank the Max Planck Society for support through the Max Planck Research Group for SHS. This research was supported in part by Perimeter Institute for Theoretical Physics. Research at Perimeter Institute is supported by the Government of Canada through the Department of Innovation, Science and Economic Development and by the Province of Ontario through the Ministry of Research, Innovation and Science.

%=====================================================================

%============================= References =============================
\bibliographystyle{yahapj}
\bibliography{mn2e}
%=====================================================================

%%=====================================================================
%% APPENDICES
\appendix
\section{Summary of models, BIC values and posterior weights}
\label{sec:appendix}
%%=====================================================================

\begin{table*}
	\caption{Extended table covering the results from our joint strong lensing and stellar dynamical models. The first column indicates the model, mock error type and S/N of the mock \jwst\ kinematics, as explained in Sections \ref{sec:data} and \ref{sec:analysis}. The second and third column displays the adopted mass model and its corresponding source resolution. The fourth and fifth column shows their respective BIC differences (with respect to the best model for a given data set and across all source models and source resolutions) and their relative posterior weights.}
	\begin{center}
	\centerline{
	\begin{tabular}{ c  c  c  c  c }
		\hline
		\hline
		Data & Model & Source Resolution & $\Delta$BIC & $f_{\mathrm{BIC}}^{*}$ \\
		\hline
		\textbf{Full FOV} & & \\
		Lensing \& Dynamics & COMPOSITE & 68 & 4.73 & 0.46 \\
		IDEAL (60/40) &  & 66 & 7.35 & 0.23 \\
		 &  & 64  & 2.00 & 0.76 \\
		 &  & 62  & 10.13 & 0.10 \\
		 &  & 60  & 11.83 & 0.05 \\
		 &  & 58  & 17.14 & 0.01 \\
		 &  & 56  & 6.69 & 0.28 \\
		 &  & 54  & 6.59 & 0.29\\
		 \\
		 & SPEMD & 68 & 128.05 & 0.0 \\
		 &  & 66 & 134.88 & 0.0 \\
		 &  & 64  & 127.72 & 0.0\\
		 &  & 62  & 135.75 & 0.0\\
		 &  & 60  & 126.76 & 0.0\\
		 &  & 58  & 126.36 & 0.0\\
		 &  & 56  & 128.98 & 0.0\\
		 &  & 54  & 146.05 & 0.0\\
		 \\
		Lensing \& Dynamics & COMPOSITE & 68 & 5.29 & 0.40 \\
		FIDUCIAL (60/40)  & & 66 & 5.74 & 0.36 \\
		 &  & 64  & 1.92 & 0.78 \\
		 &  & 62  & 13.71 & 0.02 \\
		 &  & 60  & 13.82 & 0.02 \\
		 &  & 58  & 16.89 & 0.01 \\
		 &  & 56  & 7.40 & 0.23 \\
		 &  & 54  & 6.83 & 0.28 \\
		 \\
		 & SPEMD & 68 & 134.32 & 0.0 \\
		 &  & 66  & 141.00 & 0.0 \\
		 &  & 64  & 133.18 & 0.0\\
		 &  & 62  & 142.84 & 0.0\\
		 &  & 60  & 131.63 & 0.0\\
		 &  & 58  & 131.42 & 0.0\\
		 &  & 56  & 133.70 & 0.0\\
		 &  & 54  & 148.96 & 0.0\\
		\hline
	\end{tabular}
	}
%	\vspace{2ex}
	\label{table:a1}
	\end{center}
\end{table*}

\newpage

\begin{table*}
	\caption{Extended table covering the results from our joint strong lensing and stellar dynamical models. The first column indicates the model, mock error type and S/N of the mock \jwst\ kinematics, as explained in Sec. \ref{sec:data} and \ref{sec:analysis}. The second and third column displays the adopted mass model and its corresponding source resolution. The fourth, fifth and sixth column shows their respective BIC values, BIC differences (with respect to the best model for a given data set and across all source models and source resolutions) and their relative weights.}
	\begin{center}
	\centerline{
	\begin{tabular}{ c  c  c  c  c }
		\hline
		\hline
		Data & Model & Source Resolution & $\Delta$BIC & $f_{\mathrm{BIC}}^{*}$ \\
		\hline
		\textbf{SMALL FOV} & & \\
		Lensing \& Dynamics & COMPOSITE & 68 & 3.46 & 0.63 \\
		IDEAL (60/40) &  & 66 & 7.34 & 0.29 \\
		 &  & 64  & 0.0 & 1.0 \\
		 &  & 62  & 10.33 & 0.13 \\
		 &  & 60  & 9.45 & 0.17 \\
		 &  & 58  & 11.30 & 0.10\\
		 &  & 56  & 4.36 & 0.54 \\
		 &  & 54  & 4.08 & 0.57 \\
		 \\
		 & SPEMD & 68 & 187.75 & 0.0 \\
		 &  & 66  & 181.13 & 0.0 \\
		 &  & 64  & 189.48 & 0.0 \\
		 &  & 62  & 182.40 & 0.0 \\
		 &  & 60  & 171.61 & 0.0 \\
		 &  & 58  & 182.04 & 0.0 \\
		 &  & 56  & 175.85 & 0.0 \\
		 &  & 54  & 168.12 & 0.0 \\
		 \\
		Lensing \& Dynamics & COMPOSITE & 68 & 5.66 & 0.57 \\
		FIDUCIAL (60/40) &  & 66 & 8.77 & 0.36 \\
		 &  & 64  & 0.0 & 1.0 \\
		 &  & 62  & 12.32 & 0.20 \\
		 &  & 60  & 13.67 & 0.15 \\
		 &  & 58  & 17.96 & 0.06 \\
		 &  & 56  & 6.69 & 0.49 \\
		 &  & 54  & 5.94 & 0.55 \\
		 \\
		 & SPEMD & 68 & 218.30 & 0.0 \\
		 &  & 66  & 213.94 & 0.0 \\
		 &  & 64  & 232.53 & 0.0 \\
		 &  & 62  & 214.91 & 0.0 \\
		 &  & 60  & 205.51 & 0.0 \\
		 &  & 58  & 223.16 & 0.0 \\
		 &  & 56  & 212.28 & 0.0 \\
		 &  & 54  & 194.51 & 0.0 \\
		 \hline
	\end{tabular}
	}
%	\vspace{2ex}
	\label{table:a2}
	\end{center}
\end{table*}

\newpage

\begin{table*}
	\caption{Extended table covering the results from our joint strong lensing and stellar dynamical models. The first column indicates the data and mock \jwst\ data quality, as explained in Sec. \ref{sec:data}. The second and third column displays the adopted mass model and its corresponding source resolution. The fourth, fifth and sixth column shows their respective BIC values, BIC differences (with respect to the best model for a given data set and across all source models and source resolutions) and their relative weights.}
	\begin{center}
	\centerline{
	\begin{tabular}{ c  c  c  c  c }
		\hline
		\hline
		Data & Model & Source Resolution & $\Delta$BIC & $f_{\mathrm{BIC}}^{*}$ \\
		\hline
		\textbf{PSF} & & \\
		Lensing \& Dynamics & COMPOSITE & 68 & 2.38 & 0.72 \\
		IDEAL (60/40) &  & 66 & 5.73 & 0.36 \\
		 &  & 64  & 0.0 & 1.0 \\
		 &  & 62  & 9.09 & 0.14 \\
		 &  & 60  & 10.31 & 0.09 \\
		 &  & 58  & 14.36 & 0.02 \\
		 &  & 56  & 4.63 & 0.47 \\
		 &  & 54  & 3.69 & 0.57 \\
		 \\
		 & SPEMD & 68 & 133.90 & 0.0 \\
		 &  & 66  & 139.07 & 0.0 \\
		 &  & 64  & 132.91 & 0.0 \\
		 &  & 62  & 141.86 & 0.0 \\
		 &  & 60  & 131.92 & 0.0 \\
		 &  & 58  & 131.39 & 0.0 \\
		 &  & 56  & 133.22 & 0.0 \\
		 &  & 54  & 149.98 & 0.0 \\
		 \\
		Lensing \& Dynamics & COMPOSITE & 68 & 4.07 & 0.53 \\
		FIDUCIAL (60/40) & & 66 & 4.41 & 0.49 \\
		 &  & 64  & 0.0 & 1.0 \\
		 &  & 62  & 10.32 & 0.09 \\
		 &  & 60  & 12.07 & 0.05 \\
		 &  & 58  & 15.24 & 0.01 \\
		 &  & 56  & 6.65 & 0.29 \\
		 &  & 54  & 4.91 & 0.44 \\
		 \\
		 & SPEMD & 68 & 140.07 & 0.0 \\
		 &  & 66  & 146.79 & 0.0 \\
		 &  & 64  & 138.68 & 0.0 \\
		 &  & 62  & 148.67 & 0.0 \\
		 &  & 60  & 137.23 & 0.0 \\
		 &  & 58  & 138.06 & 0.0 \\
		 &  & 56  & 139.04 & 0.0 \\
		 &  & 54  & 155.05 & 0.0 \\
		\hline
	\end{tabular}
	}
%	\vspace{2ex}
	\label{table:a3}
	\end{center}
\end{table*}

\newpage

\begin{table*}
	\caption{Extended table covering the results from our joint strong lensing and stellar dynamical models. The first column indicates the model, mock error type and S/N of the mock \jwst\ kinematics, as explained in Sec. \ref{sec:data} and \ref{sec:analysis}. The second and third column displays the adopted mass model and its corresponding source resolution. The fourth, fifth and sixth column shows their respective BIC values, BIC differences (with respect to the best model for a given data set and across all source models and source resolutions) and their relative weights.}
	\begin{center}
	\centerline{
	\begin{tabular}{ c  c  c  c  c }
		\hline
		\hline
		Data & Model & Source Resolution & $\Delta$BIC & $f_{\mathrm{BIC}}^{*}$ \\
		\hline
		\textbf{APERTURE} & & \\
		Lensing \& Dynamics & COMPOSITE & 68 & 0.00 & 1.0 \\
		IDEAL (60/40) &  & 66 & 0.03 & 0.99 \\
		 &  & 64  & 0.05 & 0.98 \\
		 &  & 62  & 0.01 & 1.00 \\
		 &  & 60  & 0.03 & 0.99 \\
		 &  & 58  & 0.00 & 1.0 \\
		 &  & 56  & 0.01 & 1.0 \\
		 &  & 54  & 0.00 & 1.0 \\
		 \\
		 & SPEMD & 68 & 0.01 & 1.0 \\
		 &  & 66  & 0.06 & 0.97 \\
		 &  & 64  & 0.03 & 0.99 \\
		 &  & 62  & 0.02 & 0.99 \\
		 &  & 60  & 0.00 & 1.0 \\
		 &  & 58  & 0.00 & 0.99 \\
		 &  & 56  & 0.06 & 0.98 \\
		 &  & 54  & 0.01 & 1.0 \\
		 \\
		Lensing \& Dynamics & COMPOSITE & 68 & 0.04 & 0.99 \\
		FIDUCIAL (60/40) & & 66 & 0.12 & 0.96 \\
		 &  & 64  & 0.08 & 0.98 \\
		 &  & 62  & 0.28 & 0.88 \\
		 &  & 60  & 0.04 & 0.99 \\
		 &  & 58  & 0.05 & 0.98 \\
		 &  & 56  & 0.03 & 0.99 \\
		 &  & 54  & 0.03 & 0.99 \\
		 \\
		 & SPEMD & 68 & 0.00 & 1.0 \\
		 &  & 66  & 0.02 & 1.0\\
		 &  & 64  & 0.01 & 1.0 \\
		 &  & 62  & 0.03 & 0.99 \\
		 &  & 60  & 0.02 & 1.0 \\
		 &  & 58  & 0.0 & 1.0 \\
		 &  & 56  & 0.12 & 0.96 \\
		 &  & 54  & 0.17 & 0.93 \\
		\hline
	\end{tabular}
	}
%	\vspace{2ex}
	\label{table:a4}
	\end{center}
\end{table*}

\begin{table*}
	\caption{Extended table covering the results from our joint strong lensing and stellar dynamical models. The first column indicates the data and mock \jwst\ data quality, as explained in Sec. \ref{sec:data}. The second and third column displays the adopted mass model and its corresponding source resolution. The fourth, fifth and sixth column shows their respective BIC values, BIC differences (with respect to the best model for a given data set and across all source models and source resolutions) and their relative weights.}
	\begin{center}
	\centerline{
	\begin{tabular}{ c  c  c  c  c }
		\hline
		\hline
		Data & Model & Source Resolution & $\Delta$BIC & $f_{\mathrm{BIC}}^{*}$ \\
		\hline
		\textbf{gNFW} & & \\
		Lensing \& Dynamics & COMPOSITE & 68 & 7.55 & 0.22 \\
		IDEAL (60/40) &  & 66 & 7.92 & 0.197 \\
		 &  & 64  & 9.93 & 0.10 \\
		 &  & 62  & 7.95 & 0.20 \\
		 &  & 60  & 7.03 & 0.26 \\
		 &  & 58  & 6.66 & 0.28 \\
		 &  & 56  & 7.35 & 0.23 \\
		 &  & 54  & 8.24 & 0.18 \\
		 \\
		Lensing \& Dynamics & COMPOSITE & 68 & 7.87 & 0.20 \\
		FIDUCIAL (60/40) & & 66 & 7.49 & 0.23 \\
		 &  & 64  & 10.25 & 0.09 \\
		 &  & 62  & 8.32 & 0.18 \\
		 &  & 60  & 7.59 & 0.21 \\
		 &  & 58  & 7.09 & 0.26 \\
		 &  & 56  & 8.13 & 0.19 \\
		 &  & 54  & 8.54 & 0.17 \\
		\hline
	\end{tabular}
	}
%	\vspace{2ex}
	\label{table:a5}
	\end{center}
\end{table*}
%\bsp

\label{lastpage}
\end{document}